\documentclass[11pt]{article}

\usepackage{lipsum}
\usepackage{graphicx}
\usepackage{xcolor}
\usepackage{amsmath}  
\usepackage{natbib}   
\setcitestyle{authoryear,open={(},close={)}} 
\usepackage[colorlinks=true,linkcolor=blue,citecolor=blue,urlcolor=blue]{hyperref}
\usepackage[french,english]{babel}
\usepackage[T1]{fontenc}
\usepackage[utf8]{inputenc}
\usepackage{minitoc}
\makeatother
\usepackage{enumitem}
\usepackage{mathtools}
\usepackage{amssymb}
\usepackage{enumitem}
\usepackage{adjustbox}
\usepackage{microtype}
\usepackage[labelfont=bf]{caption}
\usepackage{subcaption}
\usepackage{fix-cm}
\usepackage[a4paper, width=160mm, top=25mm,bottom=25mm,bindingoffset=2mm]{geometry}
\usepackage{array}
\usepackage{xfrac}
\usepackage{hypcap}
\usepackage{booktabs}
\usepackage[utf8]{inputenc}  
\usepackage{graphicx}
\usepackage{placeins}

\title{\textbf{miniBLAST: a novel experimental setup for laboratory testing of structures under blast loads}}

\author{
  Ahmad Morsel\textsuperscript{1}, 
  Filippo Masi\textsuperscript{2}, 
  Emmanuel Marché\textsuperscript{1},\\
  Guillaume Racineux\textsuperscript{1}, 
  Panagiotis Kotronis\textsuperscript{1},
  Ioannis Stefanou\textsuperscript{1},
}
\date{}
\usepackage[pagewise]{lineno}
\begin{document}
\maketitle

\vspace{-10pt}
\begin{center}
\vspace{-10pt}
\small
    $^{1}$Nantes Université, Ecole Centrale Nantes, CNRS,\\
    Institut de Recherche en Génie Civil et Mécanique (GeM), UMR 6183,\\
    Nantes, France. \vspace{3pt}\\
$^{2}$Sydney Centre in Geomechanics and Mining Materials,\\
School of Civil Engineering, The University of Sydney,\\
Sydney, Australia. 
\end{center}
\vspace{10pt}

\begin{abstract}

We present a novel experimental setup called miniBLAST, which enables systematic and repeatable laboratory tests of structures subjected to blast loads. The explosive source is based on the discharge of high electrical loads on a thin conductor, producing repeatable blast-type shock waves of controlled intensity. Conducting blast experiments under safe laboratory conditions offers significant advantages over large-scale experiments, which are expensive, require specialized personnel, are limited in number, and face repeatability issues.

In this work, we provide a comprehensive description of the setup's design rationale and technical characteristics. Moreover, we place particular emphasis on the installation phases, safety, and metrology. The explosive source is analyzed and the signature of blast shock waves is retrieved. Finally, we present an example of a masonry wall subjected to blast loads of varying intensity. We then report its dynamic response in three dimensions and at different time instances.

This new experimental setup offers a cost-effective, safe, and repeatable method to study structural dynamics under blast loads, with results that can be upscaled to real structures. It also aids in evaluating numerical models and lays the groundwork for further investigations into blast effects and mitigation.

\end{abstract}


\section{Introduction}

The increased threats against buildings and infrastructures by means of explosions (accidental and deliberate) render the study of the behavior of structures against fast-dynamic loads of paramount importance. Current knowledge of the dynamic response and failure mechanisms of structures under explosions mostly relies on numerical and analytical tools \citep{PENNETIER2015,costa2017study,masi2018study,MasiDEM}. Therefore, experiments are necessary as they significantly improve our understanding about the structural response when subjected to shock waves generated by explosions. Nevertheless, blast experiments of civil engineering structures (buildings, masonry, monuments, etc.) are limited in number when compared to other loading scenarios and, in particular, seismic loads \citep{Richardshake, DJERANMAIGRE2022100587,yadav2023shake, tomavzevivc2009seismic,makris2013planar,berto2023seismic,chaymaa2022assessment}.

Large-scale experiments of structures subjected to explosions are typically conducted in specialized testing areas with limited access \citep{godio2021experimental, pereira2015masonry, ahmad2014experimental, ngo2007blast,keys2017experimental,SIELICKI2019274}. These experiments have contributed to characterize the propagation of shock waves, the impact of different explosive sources, and of the resulting blast-induced loading \citep[to mention few]{Tyas2018, filice2022experimental, trelat2007strong, PENNETIER2015, Fouchier2017, ZYSKOWSKI2004291, Sochet19, gault2020influence}.

Without doubt, experimental testing of full scale structures exposed to blast loads is non trivial due to the associated cost, difficulties in ensuring repeatability, and of course safety risks \citep{draganic2018overview}. An alternative approach is to resort to reduced-scale experiments in laboratory conditions. Reduced-scale experiments can ensure a high degree of repeatability, low cost, and reduced safety risks. Notice that under controlled conditions, shock waves can be handily generated using small quantities of solid or chemical explosives, \citep[see][]{sochet2010blast, hargather2007optical, hargather2005determining, kleine2005laboratory, hargather2007optical,neuberger2007scaling}. 
However, reduced-scale experiments in the literature mostly focus on shock wave propagation rather than the structural response \citep{trelat2007strong, PENNETIER2015}. For example, \cite{ZYSKOWSKI2004291} studied the pressure reflection after the explosion of a gaseous mixture (hydrogen–air) in an unvented structure (closed box), and \cite{gault2020influence} examined the influence of the position of the explosive source by means of a stoichiometric propane–oxygen mixture in a confined room. To our knowledge, no experimental setups exist targeting the study of the rich dynamic behavior of structures subjected to blast loads under laboratory conditions, despite the aforementioned advantages of such tests as compared to large-scale experiments and numerical simulations. The recent advances in the derivation of scaling laws \citep{masi2020scaling} that match the physics of the explosive source with those of the structures enable the design of such a setup.

The aim of this article is to introduce a novel experimental setup, which we refer to as miniBLAST, allowing to conduct laboratory tests of structures subjected to blast loads in a systematic and repeatable manner. In minBLAST, the explosive source is an electrical system capable of generating repeatable shock waves of controlled intensity.\\

The manuscript is organized as follows. Section \ref{sec:Scaling_laws} outlines a few background elements upon which miniBLAST is built, namely the characteristics of blast loading's and the adopted scaling laws. Section \ref{sec:design} details the design rationale and the installation of the experimental setup. Section \ref{sec:exploding_wires} presents the electrical system used to generate blast loading and Section \ref{sec:system_measurements} details the data acquisition devices, sensors and metrology.
Finally, Section \ref{sec:First experimental results} showcases the capabilities of our setup by presenting experiments for a reduced-scale structure, thus illustrating the repeatability of the generated shock waves and the potential for monitoring the structural dynamics.

\section{Blast loads and scaling laws}
\label{sec:Scaling_laws}
\subsection{Blast loads}
\noindent Explosions involve rapid and sudden releases of energy, resulting in the production of large volumes of expanding gases. The impact of an explosion on structures largely depends on factors such as the size of the explosive charge, the nature of the explosive material, and the distance between the explosion and the structure.

During an explosion, a blast wave is generated due to a rapid increase in pressure, density, and temperature, which originates from the deposition of a large amount of energy in a localized volume (detonation).
The blast wave is a combination of a leading shock and a subsequent expansion zone, propagating outward from the explosive source at supersonic velocities. Across the leading shock, the thermodynamic properties of the expanding gas undergo large and abrupt changes. Within the expansion region these properties return to their initial values, albeit gradually and often in an oscillatory manner. The blast wave's propagation speed and amplitude decrease as the distance from the explosive source increases \citep[for more details, we refer to][]{sochet2010blast, ngo2007blast, mendoncca2020experimental, frost2018heterogeneous}.

\begin{figure}[htp!]
    \centering
    \includegraphics[scale=0.6]{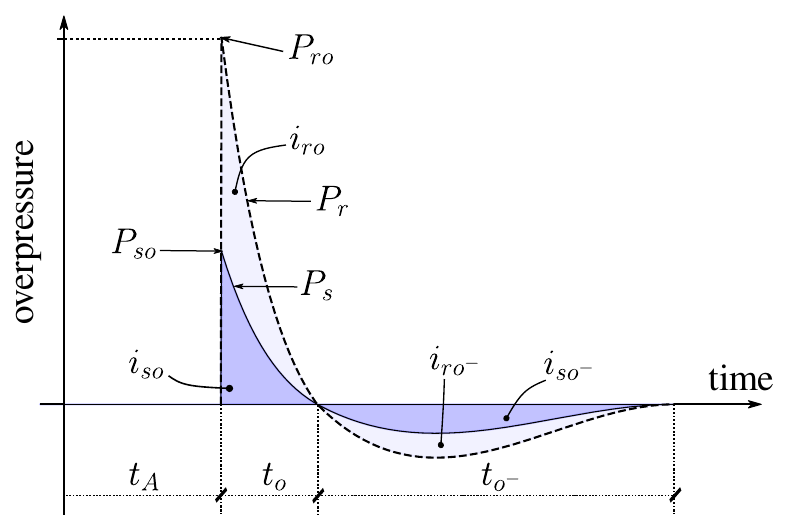}
    \caption{Time history of the incident $P_{s}$, and reflected $P_{r}$ overpressure due to an explosion at a fixed distance $D$ from the explosive source.}
    \label{fig:incident_ref_dynamic}
\end{figure}

We define hereinafter the main quantities describing the loads induced by the detonation of such explosives. We use the term overpressure to denote a differential pressure, relative to the ambient one, $P_o$. Following \cite{Dewey_blast} and Figure \ref{fig:incident_ref_dynamic}, the incident overpressure, $P_s$, is defined as the pressure solely determined by the explosion in free air (hydrostatic pressure). Conversely, the reflected overpressure, $P_r$, comprises the hydrostatic pressure increased by the reflection at the target’s surface.

Figure \ref{fig:incident_ref_dynamic} presents schematically the time evolution of the incident and reflected overpressures over time, $t$,  originating from the detonation of an explosive weight, denoted with $m$ (kg), at a distance $D$. At the arrival time, $t_A$, the incident and reflected overpressures increase with a strong discontinuity to a peak value, $P_{so}$ and $P_{ro}$, respectively. For $t > t_A$, the two overpressures decrease rapidly, with an almost exponential decay. For $t>t_A + t_o$, the overpressure further decreases to negative values and gradually approaches the ambient pressure at $t=t_A+t_o+t_{o-}$, where $t_o$ and $t_{o-}$ is the duration of the positive and negative phases, respectively.

Blast loads and their time history are often modeled relying on heuristic models \cite{friedlander1946diffraction} and best-fit interpolations of the aforementioned parameters \cite{kingery1984,vannucci2017}, expressed in terms of the scaled distance $Z=D/\sqrt[3]{m}$.

\subsection{Scaling laws}
\begin{figure}[htp!]
    \centering
    \includegraphics[scale=0.9]{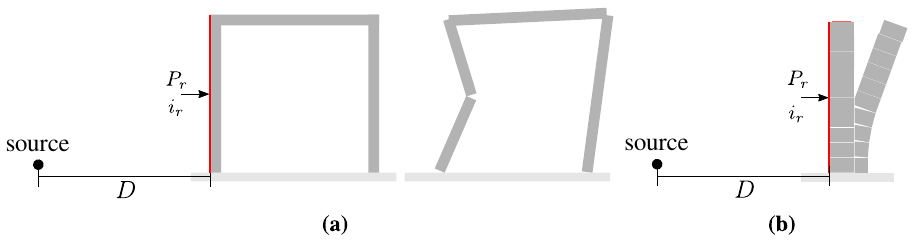}
    \caption{Representative scheme of the structural response due to blast loads considered by the scaling factors in Table \ref{table:scaling_law}: (a) a continuous structure which is transformed to a collection of interconnected blocks due to material and crack propagation and (b) masonry column made of discret building blocks.}
    \label{fig:wall_assumption}
\end{figure}

Upon the blast wave impinging on the structure, failure mechanisms due to material failure and cracks propagation are almost instantaneously generated \citep[cf.][]{masi2020resistance} leading to a system of interconnected blocks, see Figure \ref{fig:wall_assumption}. The dynamic behavior of the structure and its potential collapse will be determined by the motion of those blocks. It is, therefore, natural to adopt the scaling laws proposed by \cite{masi2020scaling}, which consider both geometric and mass scaling by means of the geometric scale factor, $\lambda = \sfrac{\tilde{l}}{l}$, and the density scale factor, $\gamma = \sfrac{\tilde{\rho}}{\rho}$. Herein, $l$ and $\rho$ denote, respectively, the characteristic length and the equivalent mass density of the full-scale structure, i.e., the prototype. The superscript tilde ($\tilde{l}$) denotes the reduced-scale structure's quantities, i.e., the model. Table \ref{table:scaling_law} summarizes the corresponding scaling factors obtained following the procedure proposed in \cite{masi2020scaling}, where $P_r$ and $i_r$ represent the total reflected overpressure and impulse acting on the structure, see Figure \ref{fig:wall_assumption}.\\

\begin{table}[htp!]
\caption{Scaling factors, see also \cite{masi2020scaling}}
\centering
 \begin{tabular}{m{10em} c m{12em} c} 
 \hline
 \small
 \textbf{Variable} & \textbf{Scaling factor} & \textbf{Variable} & \textbf{Scaling factor}\\ 
  \hline
 Length, $l$ & $\lambda$ & Linear displacement, $x$ & $\lambda$ \\ 
 Time, $t$ & $\lambda^{1/2}$ & Linear velocity, $\dot{x}$ & $\lambda^{1/2}$ \\ 
 Material density, $\rho$ & $\gamma$ & Linear acceleration, $\ddot{x}$ & 1 \\
 Mass, $m$ & $\gamma\lambda^3$ & Angle, $\theta$ & $1$ \\
 Reflected impulse, $i_r$ & $\gamma\lambda^{3/2}$ & Angular velocity, $\dot{\theta}$ & $\lambda^{-1/2}$\\
 Reflected pressure, $P_r$ & $\gamma\lambda$  & Angular acceleration, $\ddot{\theta}$ & $\lambda^{-1}$ \\
 Friction angle, $\varphi$ & $1$ & Mass moment of inertia, $J$ & $\gamma\lambda^5$\\
 \hline
\end{tabular}
\label{table:scaling_law}
\end{table}

\begin{figure}[htp!]
    \centering
    \includegraphics[scale=0.9]{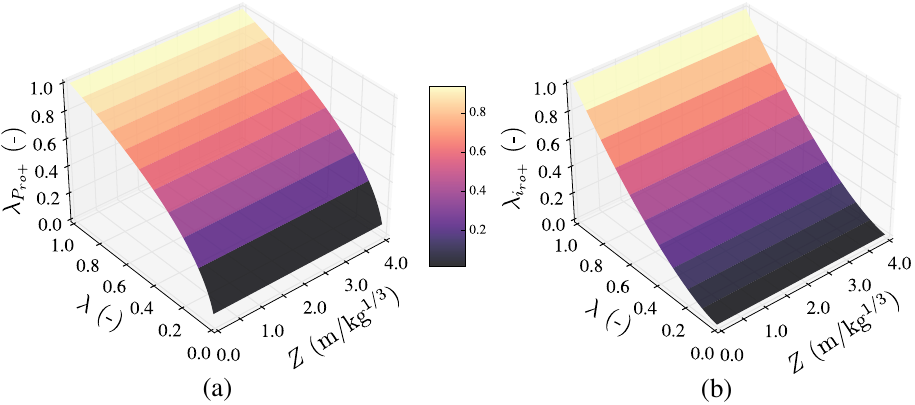}
    \caption{Scaling factors for the peak of the reflected overpressure peak $\lambda_{P_{ro+}}$ (a), and for the impulse $\lambda_{i_{ro+}}$ (b), as functions of the geometric scaling factor $\lambda$ and the density scaling factor $\gamma$.}
    \label{fig:scaling_pr_ir}
\end{figure}

Figures \ref{fig:scaling_pr_ir}(a,b) display the dependence of the scaling factors on overpressure and impulse with respect to the geometric scaling and scaled distance $Z$. Note that the scaling laws allow us to reduce the overpressure and impulse peak several orders of magnitude in function of the geometric and density scaling, $\lambda$ and $\gamma$, which demonstrates the effectiveness of the proposed scaling approach to reduce the blast load and, therefore, to enable safe tests in the laboratory. 

\clearpage
\section{Design rationale of miniBLAST}
\label{sec:design}
\begin{figure}[htp!]
    \begin{subfigure}[b]{1\textwidth}
        \centering
        \includegraphics[draft=false,scale=0.5]{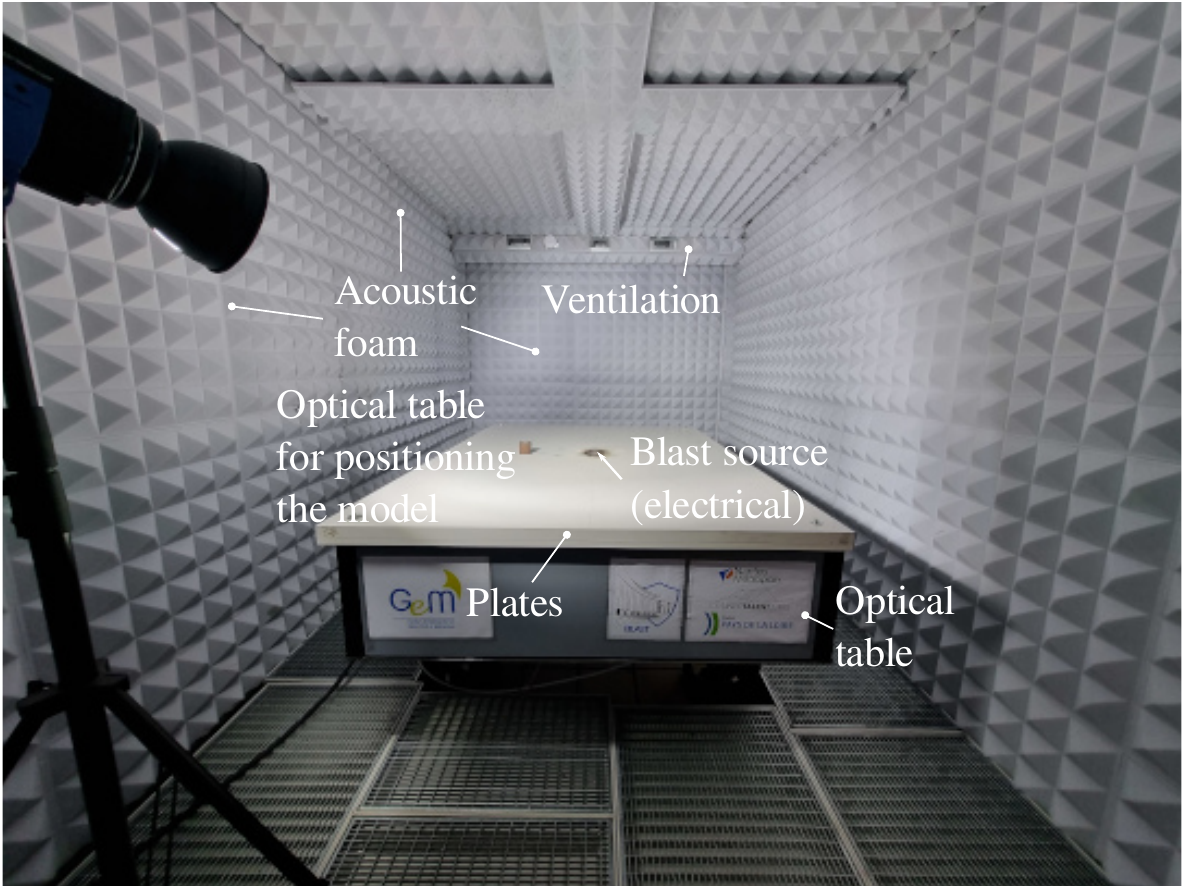}
        \caption{Inside the container cabin}
        \label{fig:room_installed}
    \end{subfigure}
    \hfill
    \begin{subfigure}[b]{1\textwidth}
        \centering
        \includegraphics[scale=0.3]{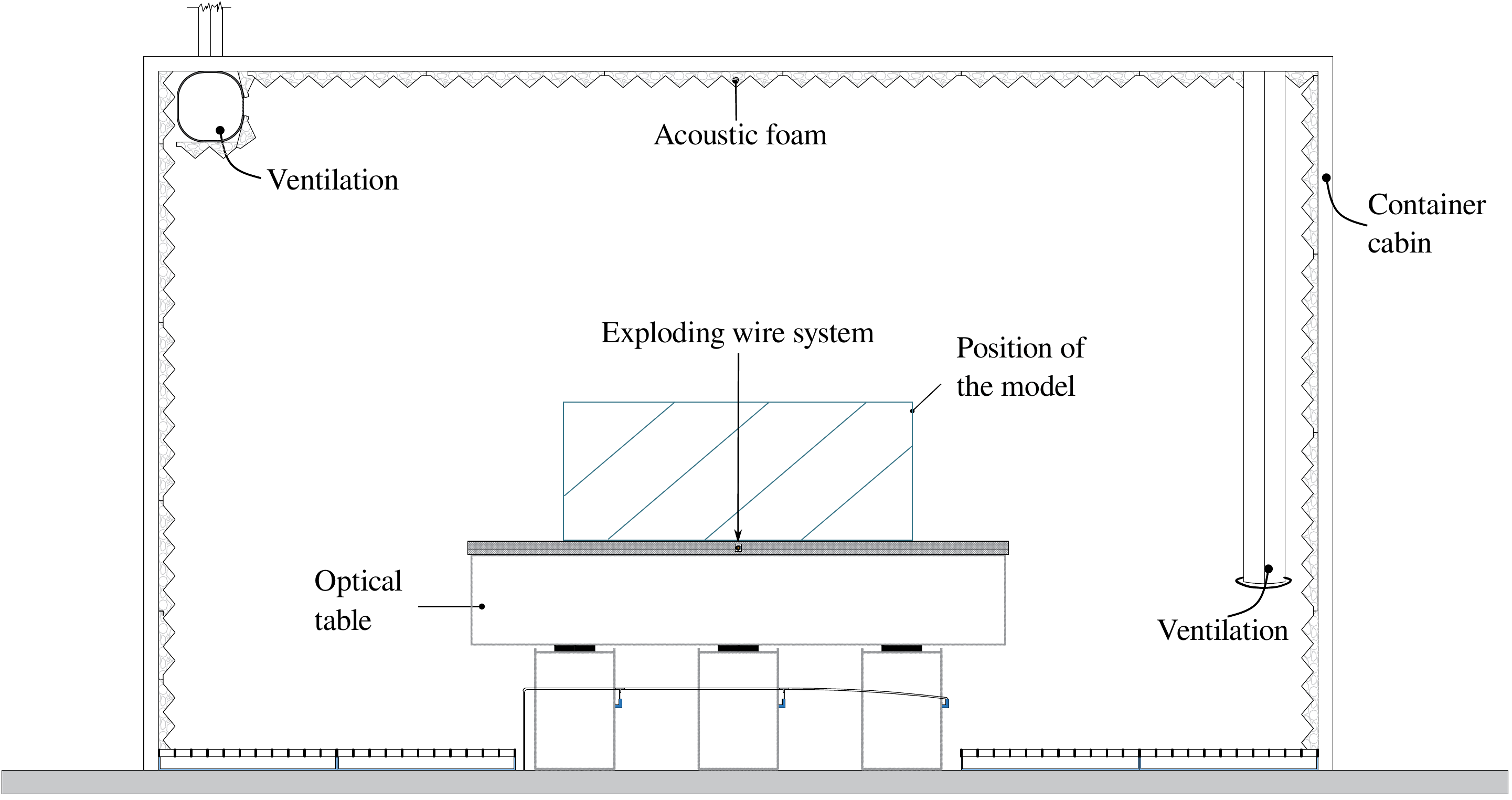}
        \caption{Longitudinal cross-section}
        \label{fig:room_drawing}
    \end{subfigure}
        \caption{Setup of miniBLAST: view from inside the container cabin (a), and longitudinal cross-section showing the components (b).}
    \label{fig:miniBLAST}
\end{figure}

\noindent  Figure \ref{fig:miniBLAST} present the setup and a schematic view of miniBLAST. The platform is designed according to the following criteria: (a) environmental safety and compliance with national and international standards for sound pressure levels \citep{french_standard, Norm_French, NIOSH_standard}, (b) repeatability and controllability of the generated blast loads, and (c) possibility to test structural models of a maximum size $120 \times 80 \times 50$ cm\textsuperscript{3} (length\texttimes width\texttimes height).

\noindent The setup comprises five main components (see Figure \ref{fig:miniBLAST}): (1) a container cabin, (2) acoustic foam siding, (3) an optical table, (4) ventilation and (5) the exploding wire system. The container cabin provides a controlled setting for carrying out experiments as it both isolates the setup from the rest of the laboratory environment and ensures the safety of individuals and of the equipment. The interior of the cabin is covered with a siding of acoustic foam. The latter absorbs and dampens the shock waves resulting from the explosion, thus minimizing eventual (secondary) reflections from the interior walls of the cabin. An optical table is positioned at the center of the cabin to provide a leveled, flat, and rigid surface where the experiments are performed, while ensuring floor isolation with the surrounding environment. Inside the cabin, a ventilation system is responsible for removing metal dust that may be produced during the detonation of exploding wires, thus providing a safe environment. Finally, the exploding wire system is our controlled explosive source and comprises two electrodes and an aluminum wire inside the container.

\subsection{Safety}
In this section we summarize the main components of the setup to assure safety during our experiments.

\subsubsection{Container cabin}
\label{subsec:Cabin}
\begin{figure}[htp!]
    \begin{subfigure}[b]{1\textwidth}
        \centering
        \includegraphics[draft=false,scale=0.5]{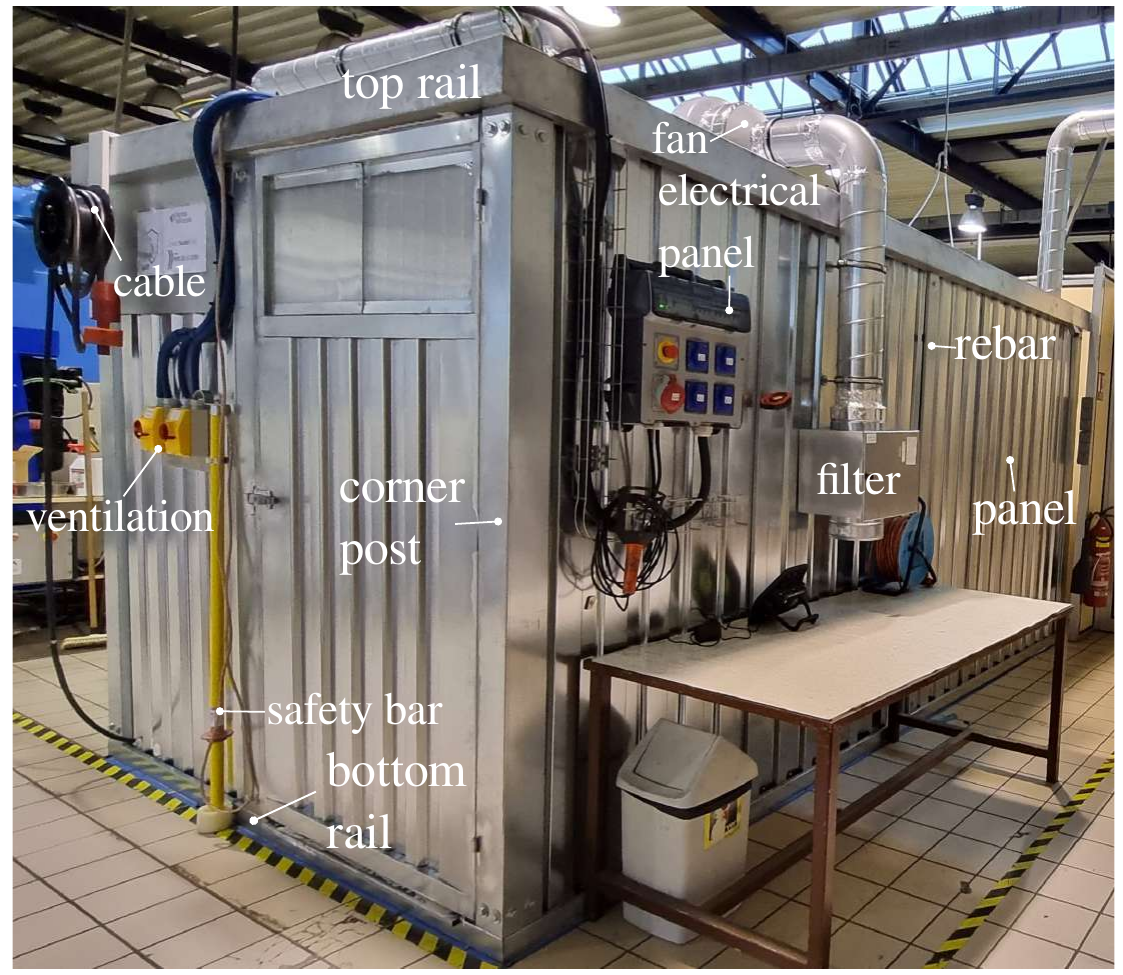}
        \caption{Container cabin}
        \label{fig:room_install}
    \end{subfigure}
    \hfill
    \begin{subfigure}[b]{1\textwidth}
        \centering
        \includegraphics[scale=0.45]{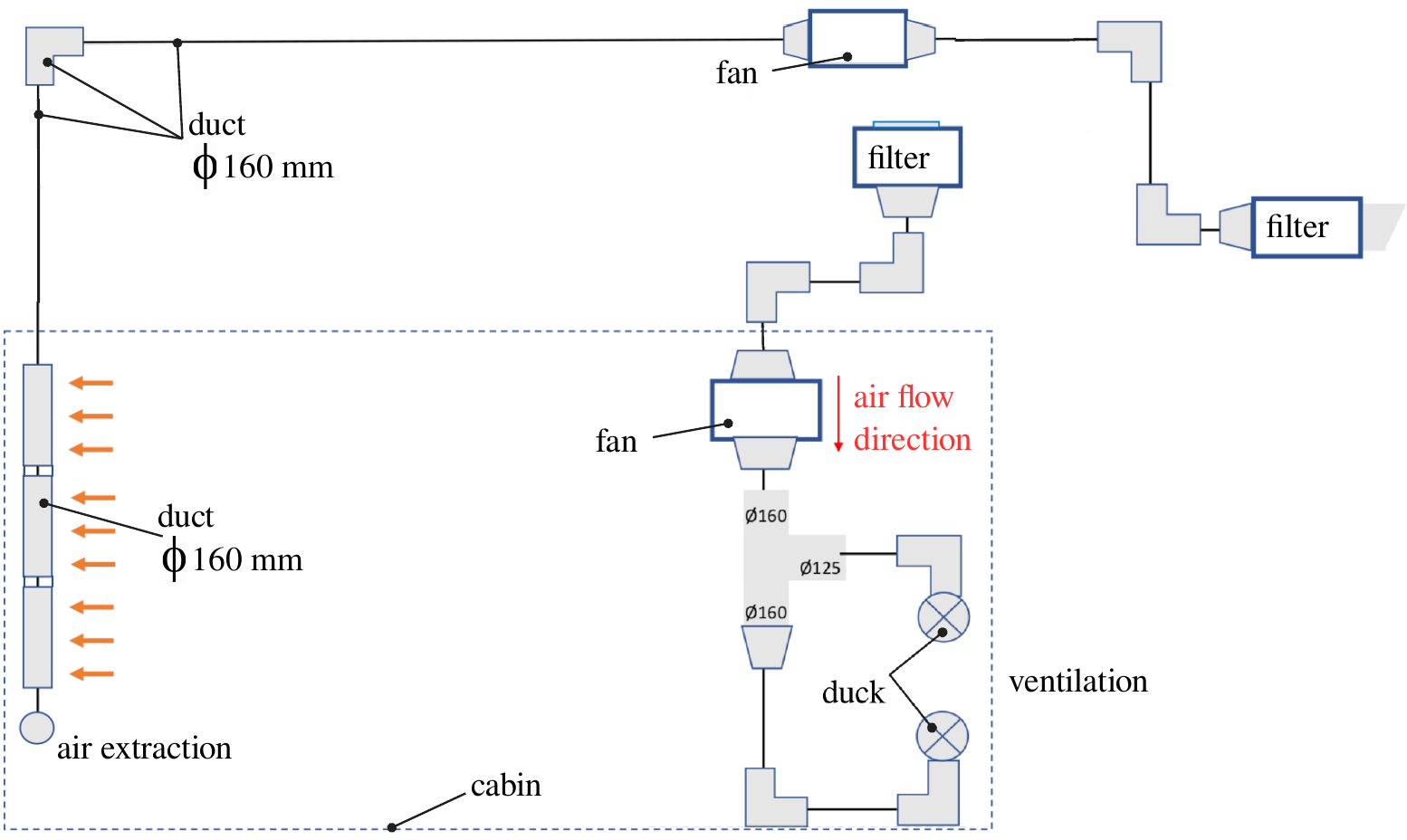}
        \caption{Ventilation system}
        \label{fig:ventilation}
    \end{subfigure}
        \caption{Setup of the ventilation system: Galvanized steel container cabin of length $4$ m, width $2.3$ m, and height $2.2$ m (a), and ventilation system installation details (b).}
    \label{fig:ventilation_system}
\end{figure}

\noindent A galvanized steel cabin \citep{room} was selected and installed. The cabin has the following dimensions: $4\times 2.3 \times 2.2$ m\textsuperscript{3} (length \texttimes width \texttimes height). This container allows to test relatively large models, while containing the blast and the generated micro-particles (see Section \ref{sec:Physical_processes}). The container is adequately reinforced with re-bars as shown in Figure \ref{fig:room_install}. Elastic bearing and fasteners were used to dissipate high frequency oscillations due to the blast loads. For more details we refer to \cite{morsel2024fast}.

\subsubsection{Ventilation system}

A ventilation system is necessary to remove metal dust particles produced during the detonation (see Section \ref{sec:exploding_wires} for detail about the electrical blast) and to fully replace the cabin's air (approximately 20 m$^3$) within a specific time-frame (2 minutes). The fan and duct connections expel the air outside the lab, using a filter for environmental compliance (Figure \ref{fig:ventilation}). In prioritizing safety, we incorporate an additional phase that combines ventilation and air extraction to ensure renewal of the cabin air. This phase necessitates another fan and a filter to regulate incoming air and protect the fan's operation. For more details see \cite{morsel2024fast}.

\subsubsection{Sound level}
\label{subsec:safety}
Based on the scaling laws and the range of pressures that can be generated with the exploding wire setup, attention is focused on the assessment of the requirements necessary to prevent any hearing loss or injury. In doing so, we define the sound pressure level $L_s$ (in dB) generated during the explosion as
\begin{equation}
    L_s = 20\log_{10}\left(\frac{P_{so}}{P_o}\right),
\label{eq:scale_sound}
\end{equation}
where $P_o=20$ \textmu Pa is the reference atmospheric pressure and $P_{so}$ is the peak of the blast incident overpressure at a distance $D$ from the explosive source. We compute $P_{so}$ relying on available best-fit approximations \citep[see][]{vannucci2017,masi2020fast} as function of the scaled distance $Z$.

\begin{figure}[htp!]
    \centering
    \includegraphics[scale=0.55]{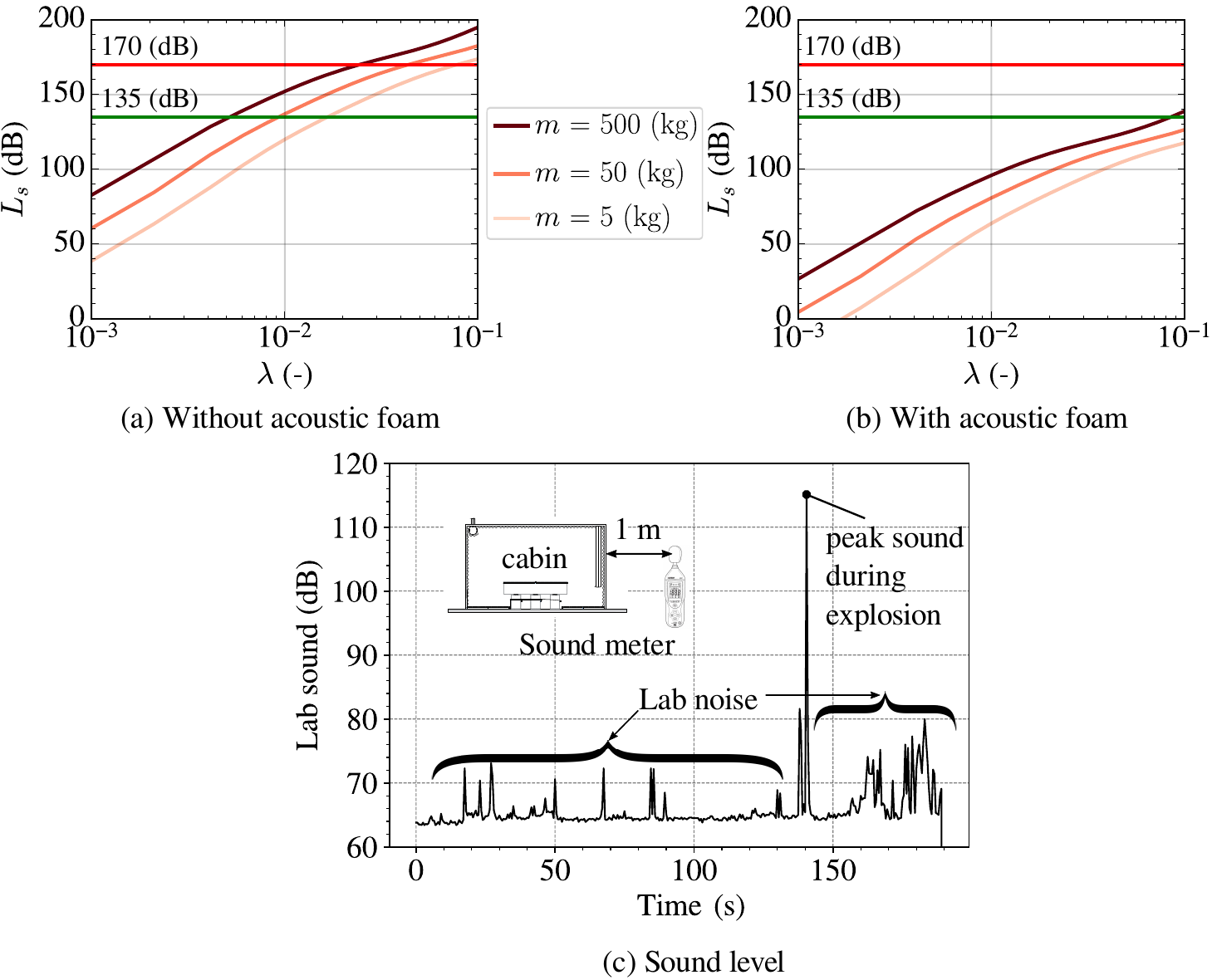}
    \caption{Sound pressure level: computed relying on best-fit interpolations at a distance equal to 10 from the cabin (a-b) and measured with a sound meter at 1 m from the cabin (c). The computed sound level is a function of $\lambda$ for different $m$ (mass of TNT for the prototype), either neglecting the acoustic foam (a) or accounting for it (b).}
    \label{fig:safety}
\end{figure}

To evaluate the sound level, we consider, for the prototype, an explosive mass $m$ of TNT explosive equivalent to that of the exploding wire employed here, for the model. Using scaling factors, $\gamma$ and $\lambda$, we compute the mass at the reduced scale, and from this mass we calculate the incident overpressure at a distance of 1.1 m (the distance between the explosive source and the cabin wall). Based on this pressure, we compute the sound level with and without the acoustic foam, as presented in Figure \ref{fig:safety}(a,b), at a distance from the cabin equal to 10 m, using the inverse distance proportional law.

Figure \ref{fig:safety}(a,b) presents the sound level $L_s$ at varying of the equivalent explosive mass, $m$, for different values of the geometric scaling factor, $\lambda$. The green line represents the maximum authorized sound level (135 dB) according to the French standards \citep{french_standard}. Moreover, using earmuff allow us to attenuate 35 dB more which correspond to a limit of 170 dB sound level (red line). We observe that the experiments exceed the acceptable sound level of $135$ dB for certain values of $\lambda$ without accounting for the acoustic foam siding. On the contrary,  the acoustic foam siding enables to dampen and reduce the sound level to safe values, cf. Figure \ref{fig:safety}(b).

To verify the above findings, we also conduct real-time measurements with a sound meter for the largest value of the overpressure achievable with our setup (cf. Section \ref{sec:exploding_wires}). The measured sound pressure levels are found to be below the maximum sound level of $135$ dB. Figure \ref{fig:safety}(c) shows that the sound levels consistently fall within the permissible range, with a maximum peak value of $115.1$ dB recorded at a distance of 1 m from the cabin.

\subsubsection{Acoustic foam siding for sound isolation and reflection avoidance}
\label{subsec:acoustic}

\noindent Acoustic foam was installed inside the container as shown in Figure \ref{fig:room_installed}. The foam absorbs and dampens the shock waves from the explosion and prevents reflections which can effect our experiments. The selected acoustic foam \citep[SE40M1][]{Foam_Solutions}, shown in Figure \ref{fig:room_installed}, is a fire-resistant melamine foam, with a pyramid shape and dimensions equal to $60 \times 60 \times 5.5$ cm\textsuperscript{3} (length\texttimes width\texttimes thickness). The acoustic foam is installed by cleaning and marking the walls and roof, applying adhesive to the panel regions, and, when needed, cutting the panels into smaller pieces for an accurate installation at corners and edges, cf. Figure \ref{fig:room_installed}. \\

\begin{figure}[htp!]
    \begin{subfigure}[b]{1\textwidth}
        \centering
        \includegraphics[scale=0.5]{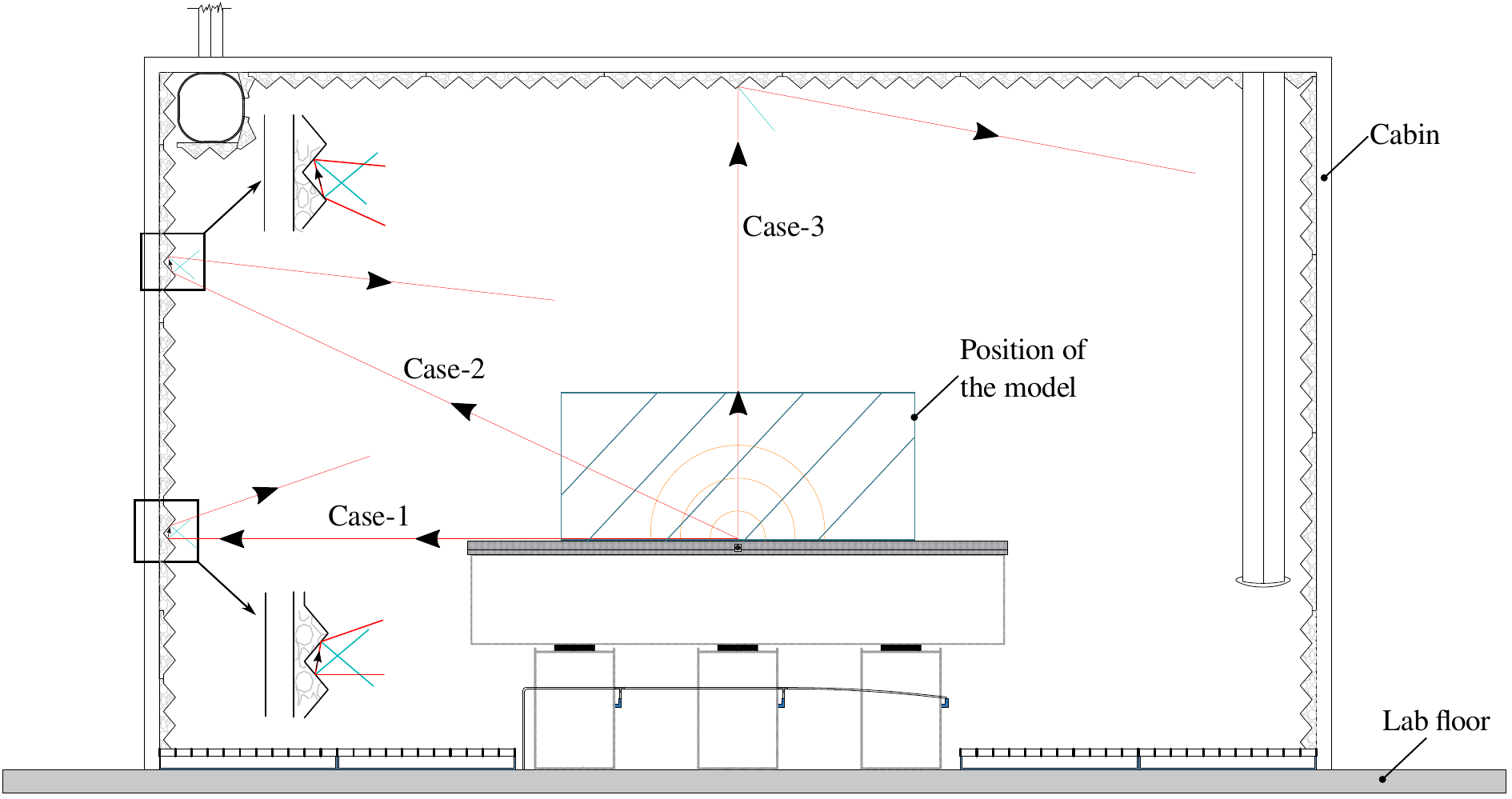}
        \caption{Longitudinal cross-section}
        \label{fig:foam_refl_long}
    \end{subfigure}
    \hfill
    \begin{subfigure}[b]{1\textwidth}
        \centering
        \includegraphics[scale=0.5]{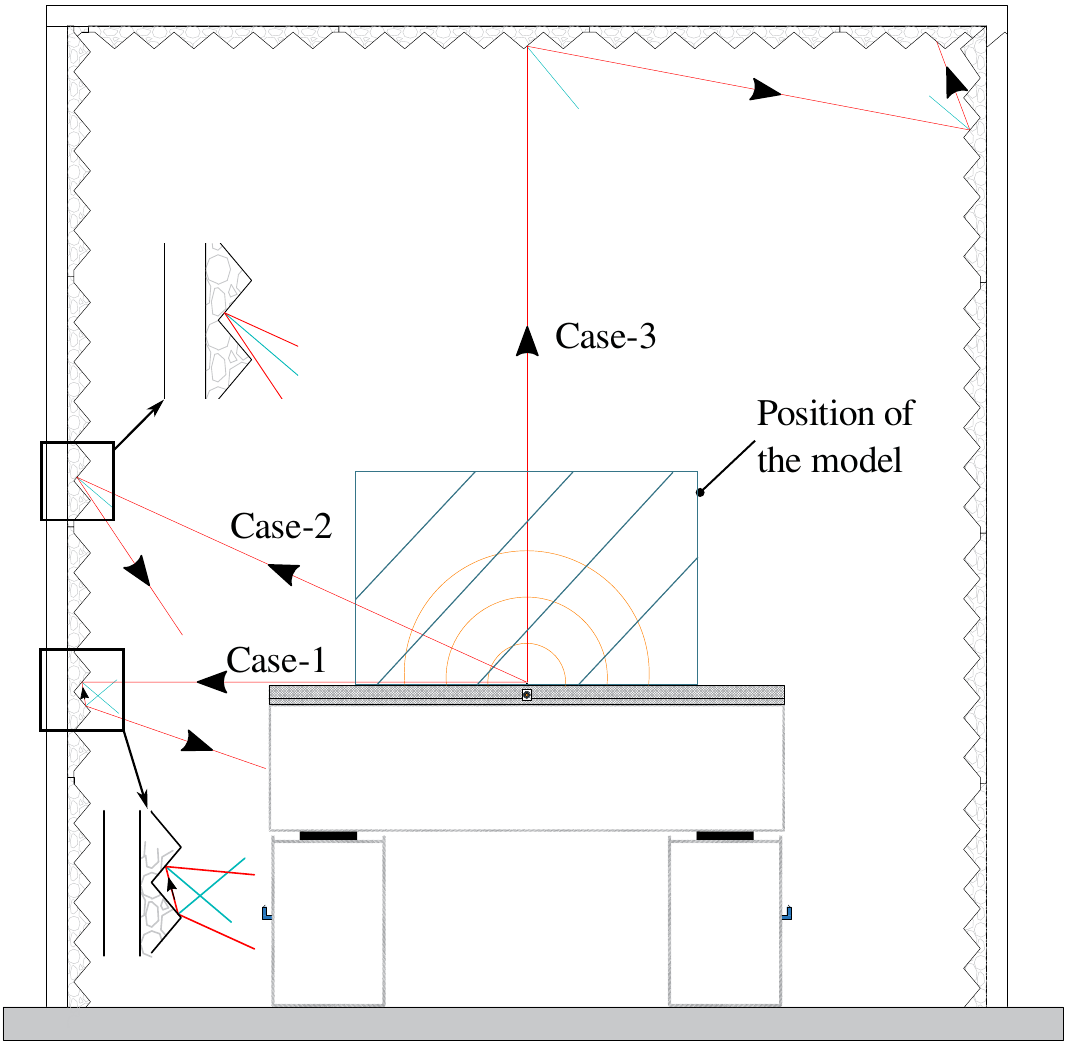}
        \caption{Transverse cross-section }
        \label{fig:foam_refl_trav}
    \end{subfigure}
        \caption{Shock wave reflections according to the position of the model: longitudinal (a) and transverse (b) cross section.}
    \label{fig:foam_reflection}
\end{figure}

As far as it concerns the reflections of the shock wave from the container wall, we considered three critical cases for selecting the shape of the panel. In Figure \ref{fig:foam_reflection}, we trace the shock wave propagation to ensure that it is reflected away from the location of the model.

\begin{figure}[htp!]
    \centering
    \includegraphics[scale=0.65]{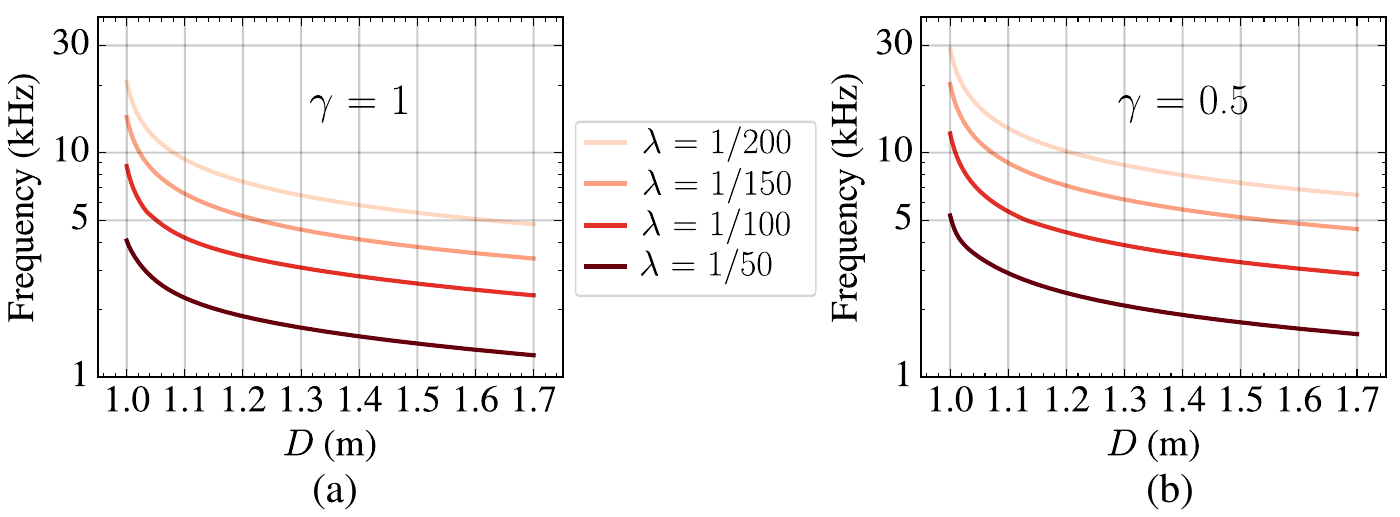}
    \caption{Shock wave frequency as a function of the standoff distance $D$, for $\gamma  = 1.0$ (a), and  $\gamma = 0.5$ (b), $\lambda \in [1/50, 1/100, 1/150, 1/200]$, for a prototype's explosive mass equal to $500$ kg.}
    \label{fig:frequency}
\end{figure}

The sound reduction values are measured for a 31 mm thick flat plate (1 mm thick metal plate plus 30 mm foam). The resulting reduction is equal to 17.21 dB for frequencies between 0.1 and 0.4 kHz, 37.53 dB for frequencies between 0.4 and 2.0 kHz, and 49.01 dB for frequencies between 2.0 and 10.0 kHz.\\
In our setup, we opt for a a foam plate with a higher thickness, namely 55 mm, which offers even higher sound attenuation. However, the exact specifications for such thickness are not available, hence, as it follows, we consider the sound reduction values for the nominal thickness of 30 mm. To check the sound level reduction in our experiments, we estimate the shock wave frequency $f_r$ as the inverse of the positive time duration of the shock wave $t_o$, i.e., $f_r = 1/t_o$. Figure \ref{fig:frequency} displays the variation of the shock wave frequency (in kHz) as a function of the standoff distance $D$. Consequently in our experiments, the frequency of the shock wave ranges between 1.2 kHz and 28.9 kHz and the anticipated acoustic foam sound reduction is greater than 37.53 dB.

\subsection{Model's support - optical table}
\label{subsec:Non-magnetic optical table}
\noindent A (MKS Newport) non-magnetic optical table, shown in Figure \ref{fig:room_installed}, is selected to ensure floor isolation with respect to the laboratory environment. The table is stiff enough to endure the load coming from the explosion without undergoing notable deflection and vibrations that could influence the response of the structure and the sensors (see Section \ref{subsec:sensor}). The table is composed of six pneumatic supports with an alignment accuracy of $\pm$ 0.3 mm, and a honeycomb stiff panel $120$ cm wide, $180$ cm long, and $30.5$ cm high, weighting $500$ kg. The table is positioned in the middle of the container cabin with a minimum spacing of 40 cm between the table and the cabin wall to allow passage and delay the dampened reflections.\\

The deflection of the table panel is estimated to $1.50$ (\textmu m) for a concentrated mass of $150$ kg at its center. This mass corresponds to a pressure of 46 kPa applied to a 20 cm diameter area at the center of the table, and represents the maximum design load considered in our design due to blast source.

The table supports comprise a system of stabilizers to isolate the table from eventual floor vibrations.

\subsection{Electrical explosive source}
\label{sec:exploding_system_design}

\noindent Exploding wires are used to safely generate blast loads (at reduced scale) in the miniBLAST controlled laboratory environment. The exact mechanism for the generation of the blast waves is presented in the next section. Here we describe succinctly the technical details of the setup and the installation.

\begin{figure}[htp!]
    \begin{subfigure}[b]{1\textwidth}
        \centering
        \includegraphics[scale=0.5]{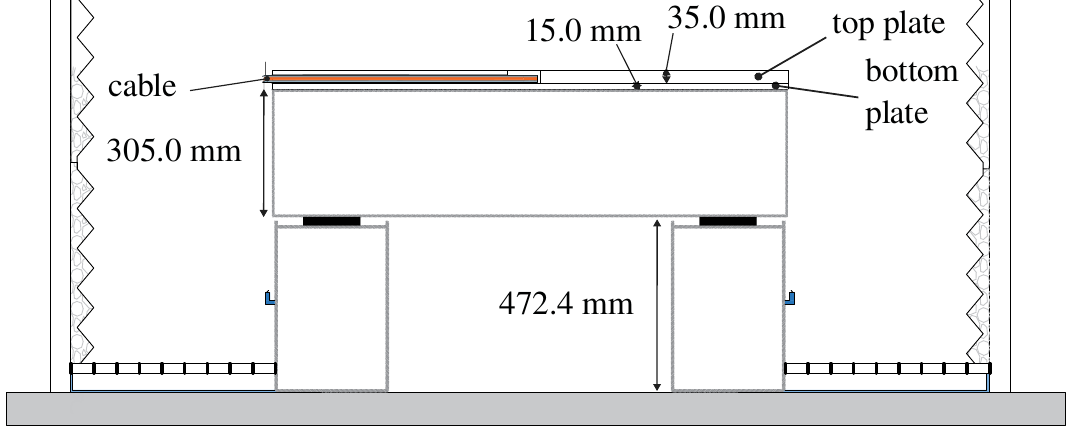}
        \caption{System's cross section}
        \label{fig:wire_system}
    \end{subfigure}
    \hfill
    \begin{subfigure}[b]{1\textwidth}
        \centering
        \includegraphics[scale=0.5]{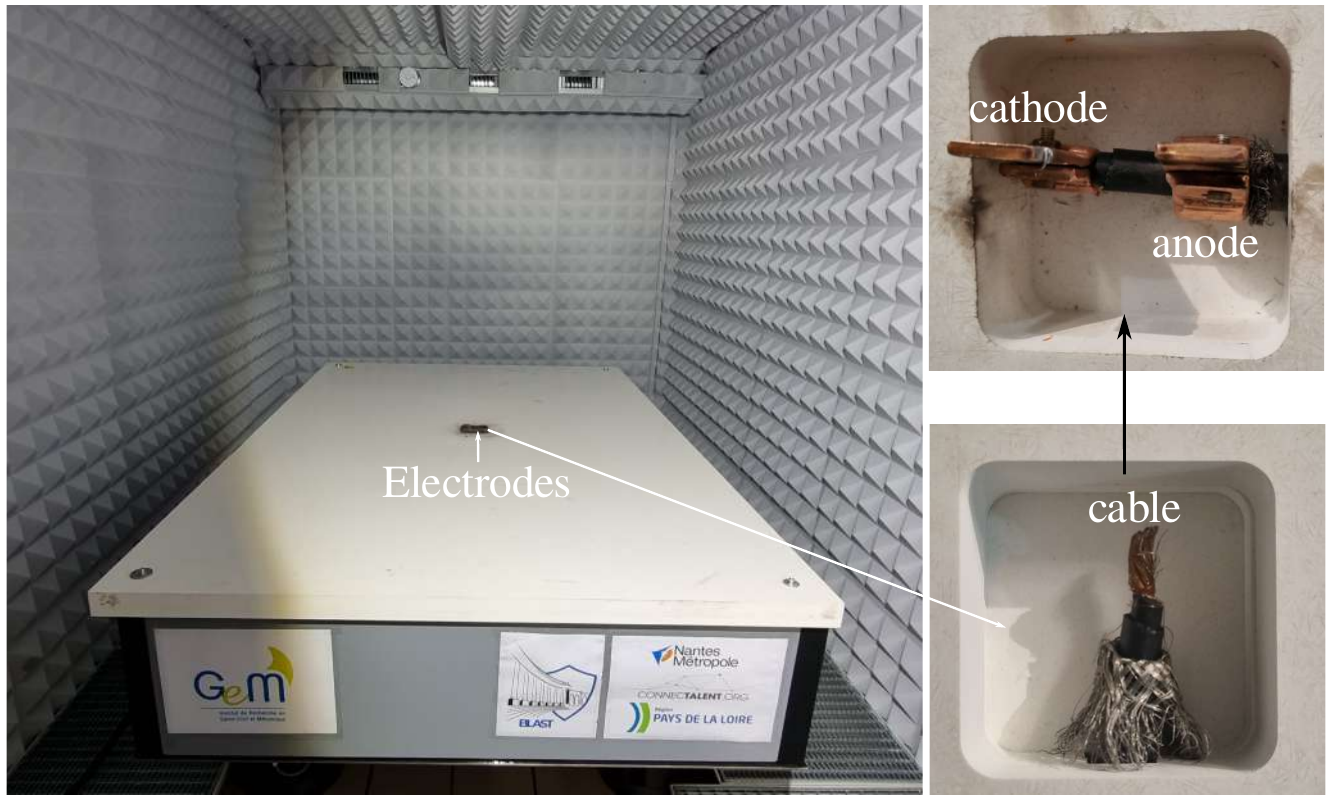}
        \caption{Installation}
        \label{fig:room_plate}
    \end{subfigure}
        \caption{Exploding wire system: (a) cross section and (b) installation of the cable and of the electrodes.}
    \label{fig:electric_wire}
\end{figure}

The exploding wire setup main components are a capacitor, a cable, a wire and two electrodes. The capacitor is connected to the electrodes via a $16$ mm thick cable that traverses the container cabin, while the cable head is positioned at the center of the optical table, Figure \ref{fig:wire_system}. The two electrodes (a cathode and an anode) are positioned on the cable's head, Figure \ref{fig:room_plate}. The cable passes through two white GPO3 Polyester glass mat plates (bottom plate thickness 1.5 cm and top plate thickness $3.5$ cm) positioned on the table. In that way, not only the cable is isolated from the table's surface, providing a safe and isolated path, but also only the electrodes are visible on the table's surface so as to be able to create (nearly) hemispherical explosions.

\newpage

\section{Controlled generation of shock waves}
\label{sec:exploding_wires}

\noindent An electrical circuit is employed to generate shock waves that simulate the effects of an explosive detonation. The circuit is connected to a thin metallic wire, which explodes when thousands of amperes pass through it. Herein, we illustrate the main physical phenomena occurring during the aforementioned process of shock generation. Next, we determine the time varying, nonlinear evolution of the resistance of the wire and we present estimations of its temperature rise and of the energy deposition within the system. All these measurements and derivations are of utmost importance as they form the basis for understanding the resulting blast loading, its intensity and repeatability, and therefore the consequent structural response.

\subsection{Charge and discharge circuits}
\label{sec: Charge and discharge circuits}

\noindent Experiments are carried out with a Pulse Current Generator (PCG) which, in simple terms, contains several large capacitors and allows to develop high currents passing through the exploding wire.

The PCG consists of two circuits, as shown schematically in Figure \ref{fig:exploding_mechanism}(a): (1) a charging circuit composed of an electric source, a high current switch, and twelve capacitors, and (2) a discharge circuit comprising six Ignitron switches (Ignitron NL7703 mercury switches) and a coaxial cable that connects the capacitor with the electrodes. For every two capacitors, one NL7703 Ignitron switch is used to control the discharge. By varying the number of the connected capacitors, the characteristics of the PCG can be changed in terms of its resistance, capacitance, and inductance.

The twelve capacitors are connected in parallel and have an equivalent nominal capacitance $C=408$ \textmu F. The resistance of the discharge circuit, $R$, that connects the PCG to the exploding wire through electrodes results from the internal resistance of the PCG and the resistance of the coaxial cable which is 10.7 m long. In a similar way, the total inductance of the discharge circuit, $L$, is due to the internal inductance of the PCG and that of the coaxial cable.

The capacitor can be charged to a maximum voltage of $15$ kV. The NL7703 Ignitron switches can handle a maximum voltage equal to $20$ kV, with peak currents of $100$ kA. The energy that can be stored within the capacitor is in between $5$ J and $46$ kJ. 

\subsection{Physical processes}
\label{sec:Physical_processes}

\begin{figure}[htp!]
    \centering
    \includegraphics[scale=0.35]{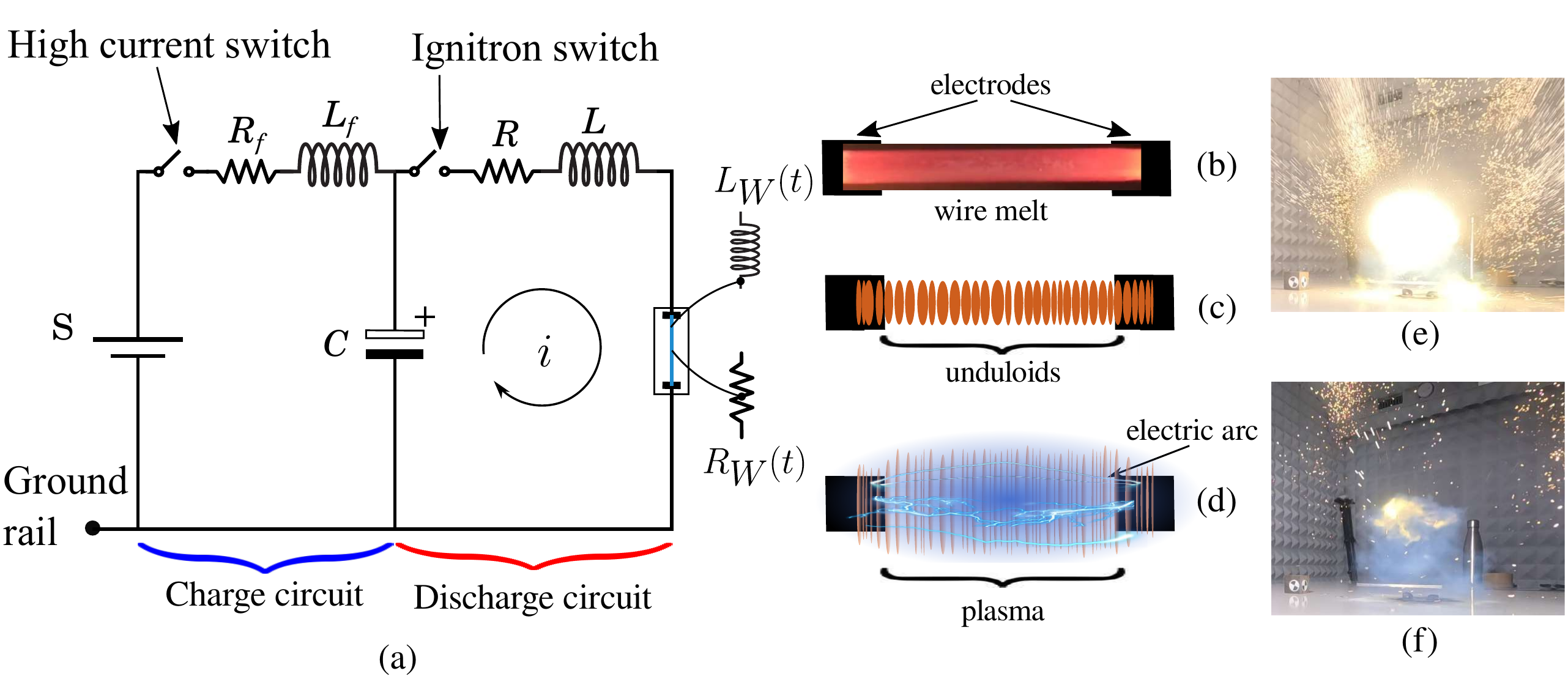}
    \caption{Schematic representation of the PCG charge and discharge circuits (a) and stages of the explosion mechanism: (b) wire melt, (c) formation of unduloids, (d) electrical breakdown and the appearance of electric arcs (plasma), (e) explosion of a wire producing a spherical light at time instance $t = 12.5$ ms, and (f) dust produced at $t = 25$ ms. Images from the miniBLAST platform using a GoPro11 camera with 240 fps and a resolution of 2704 $\times$ 1520 pixels.}
    \label{fig:exploding_mechanism}
\end{figure}

\noindent A typical exploding wire system, as the one in Figure \ref{fig:exploding_mechanism}(a), comprises a discharge circuit that incorporates a capacitor, denoted as $C$, a thin conductor (e.g. copper, tungsten, gold, aluminum, etc.) of electrical resistance $R_W$ and inductance $L_W$, a resistor $R$, an inductor $L$, and a switch. When a sufficiently large electric charge is discharged through the thin metallic conductor, its temperature rapidly increases. This fast temperature increase causes the conductor to undergo phase transitions and to generate a pressure shock wave accompanied by the emission of a bright flash of light and smoke. The whole process resembles to natural lightning but in a smaller scale, see Figure \ref{fig:exploding_mechanism}(e,f). According to \citet{Kenneth1972_wire}, several phenomena take place in the course of the following stages:

\begin{enumerate}[label= S\arabic*.]
    \item During discharge, the current heats up the conductor (ohmic heating) until it transitions to a liquid state (melting), as shown in Figure \ref{fig:exploding_mechanism}(b).
    \item Upon reaching the melting point, the current continues to flow through the thin conductor, causing the temperature to increase further. As a result, the conductor expands in volume and transitions from a liquid to a gas state -- known as the boiling stage. However, this transition is not uniform across the conductor's length. Experimental results \citep{liu2019experimental} show the formation of onduloids (or striations), as depicted schematically in Figure \ref{fig:exploding_mechanism}(c). At this stage, the apparent electrical resistance of the wire significantly increases, as the air between the onduloids acts as an insulator.
    \item The electrical (dielectric) breakdown of the air between the onduloids occurs, hence the conductivity momentarily increases, see Figure \ref{fig:exploding_mechanism}(d). This stage is characterized by the appearance of electric arcs, i.e., of plasma, marking the final phase transition in the system. The plasma rapidly expands in space due to its electrical conductivity.
\end{enumerate}
\noindent The last stage is accompanied with the generation of a pressure shock wave that propagates outwards the conductor and towards the structure.

In our case we use an aluminum wire as the conductor. The wire's diameter and the current density control the amount of energy dissipated into it during the aforementioned process and the rate at which the plasma expands \citep{Tucker_Neilson_1959, liu2019experimental}. According to \cite{zhao2013plasma}, the greater the amount of dissipated energy into the wire until the voltage peak at the end of stage S2, due to the increased resistivity, the higher the rate of expansion of the plasma. In addition, thinner wires result in a more uniform expansion of the plasma along the wire's length, leading to higher plasma expansion rates \citep{chandler2002relationship}. The wire's material also affects the plasma dynamics. Aluminum exhibits a relatively higher resistivity than copper, resulting in higher energy dissipation.  

\subsection{Aluminum wire's characteristics}
\label{Aluminium wire's characteristics}

\subsubsection{Dissipated energy}
\label{Dissipated energy}

\noindent The specific heat at constant strain is defined as follows
\begin{equation}
    C_V = \frac{dq_W (T)}{dT},
\label{eq:dQ}
\end{equation}
where $q_W$ is the heat mass density at a point in the wire and $T$ is the temperature. In the following, we assume uniform conditions over the entire mass of the wire for the sake of simplicity.

The heat, $Q_w$, required to change the temperature of the whole wire by $\Delta T=T_2-T_1$ is equal to:
\begin{equation}
    \Delta Q_W = \int_{m_W} \int_0^{\Delta T} \frac{dq_W(T)}{dT} \;dT = \int_{m_W} \int_0^{\Delta T} C_V \;dT = C_V m_W \Delta T,
\label{eq:dQ2}
\end{equation}
\noindent where $\Delta Q_W = Q_W(T_2) - Q_W (T_1)$ and $m_W$ is the mass of the wire. 

According to the laws of thermodynamics and assuming that a part of the electric energy, $\Delta E_R$, flowing through the wire is transformed to heat (ohmic heating), the energy dissipated through the wire can be calculated as
\begin{equation}
    \Delta Q_W = \chi \Delta E_R,
\label{eq:dE}
\end{equation}
\noindent where $0\leq\chi\leq 1$ is the performance coefficient. For $\chi = 1$, the dissipated energy though the wire can be calculated as
\begin{equation}
    \Delta E_R = \Delta {T} m_W C_V.
\label{eq:dEr}
\end{equation}

\noindent The nominal thermal properties of aluminum wires are summarized in Table \ref{table:thermal_prop_AL}. 
\begin{table}[h!]
\caption{Thermal properties of aluminum wires.}
\centering
\footnotesize
 \begin{tabular}{l l l l} 
 \midrule
 \textbf{Parameters} & \textbf{Value} && \textbf{Reference}\\
 \midrule
Melting temperature, $T_{melt}$ & 933& K & \citet{yang2015experimental} \\
Boiling temperature, $T_{boil}$ & 2467& K & \citet{Leitner2017} \\
Heat capacity of liquid, $C_{V-\mathrm{liquid}}$ & 1.18 &Jg$^{-1}$K $^{-1}$ & \citet{yang2015experimental} \\
Heat capacity at $T$ = 24$^\circ$ & 0.9 &Jg$^{-1}$ & \citet{yang2015experimental} \\
 \midrule
\end{tabular}
\label{table:thermal_prop_AL}
\end{table}

\begin{figure}[ht]
     \centering
     \begin{subfigure}[b]{0.47\textwidth}
         \centering
         \includegraphics[scale=0.5]{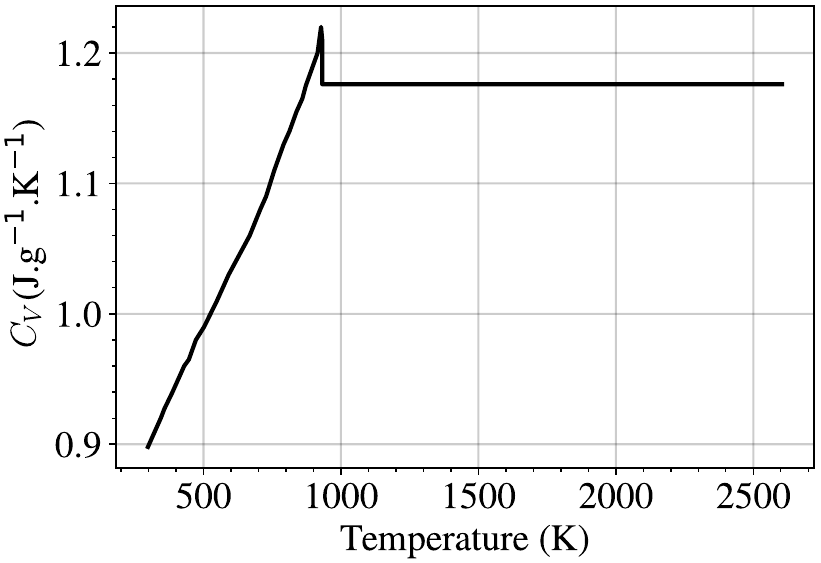}
         \caption{Heat capacity}
         \label{fig:CP}
     \end{subfigure}
     \hfill
     \begin{subfigure}[b]{0.47\textwidth}
         \centering
         \includegraphics[scale=0.5]{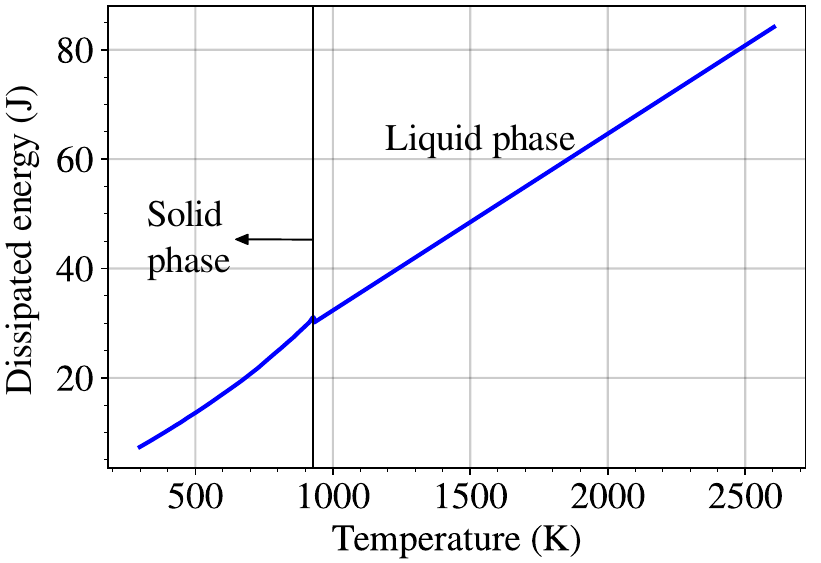}
         \caption{Dissipated energy}
         \label{fig:dEr}
     \end{subfigure}
        \caption{Specific heat per unit mass with respect to temperature (digitized from \citet{liu2019experimental}) (a) and energy dissipated through the wire as function of temperature (b) for an aluminum wire of length equal to 3.6 cm and diameter equal to 0.6 cm.}
        \label{fig:capacity_energy}
\end{figure}

\noindent The specific heat $C_V$ varies in time due to the phase transitions that occur in the aluminum wire (Section \ref{sec:Physical_processes}). For simplicity, only two phases are considered hereafter (solid and liquid, see \cite{desai1984electrical, liu2019experimental}). The values of $C_V$ as a function of temperature are digitized from \citet{liu2019experimental} (see also \citet{desai1984electrical}) and are presented in Figure \ref{fig:CP}. Using Eq. (\ref{eq:dEr}), we estimate the energy dissipated in the liquid and solid phases under normal laboratory conditions ($T = 24^\circ$C, humidity: 50$\%$) for the wires employed in our setup, that have a diameter of $0.6$ mm, length $3.6$ cm, and density $2.71$ g/cm$^3$, see Figure \ref{fig:dEr}.
 
\subsubsection{Resistance}
\label{subsec: Resistance}

\begin{figure}[ht]
     \centering
     \begin{subfigure}[b]{0.47\textwidth}
         \centering
         \includegraphics[width=0.95\textwidth]{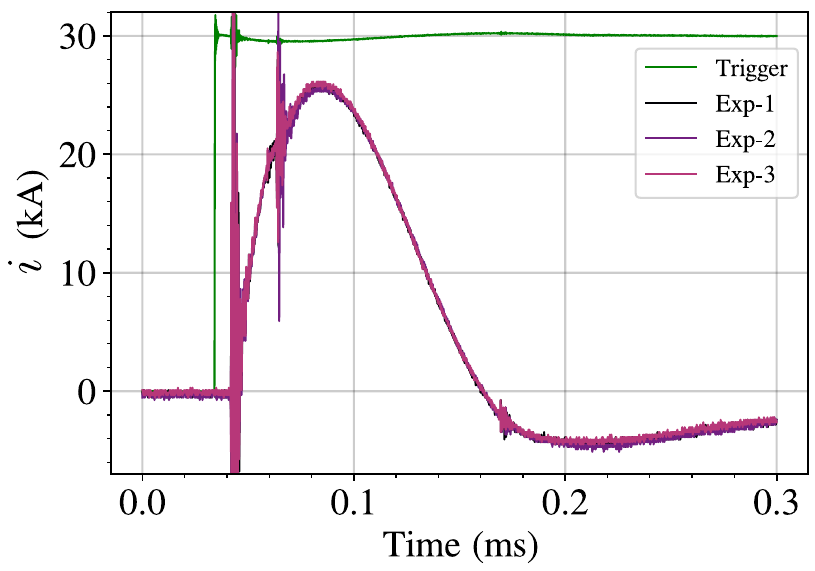}
         \caption{Discharge current}
         \label{fig:current_w}
     \end{subfigure}
     \hfill
     \begin{subfigure}[b]{0.47\textwidth}
         \centering
         \includegraphics[width=0.95\textwidth]{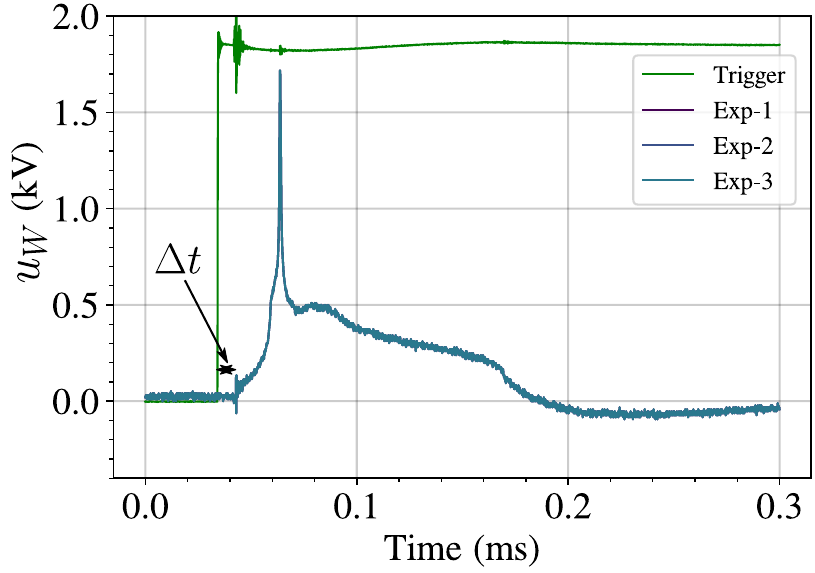}
         \caption{Discharge voltage}
         \label{fig:volt_w}
     \end{subfigure}
     \hfill
     \begin{subfigure}[b]{0.47\textwidth}
         \centering
         \includegraphics[width=0.95\textwidth]{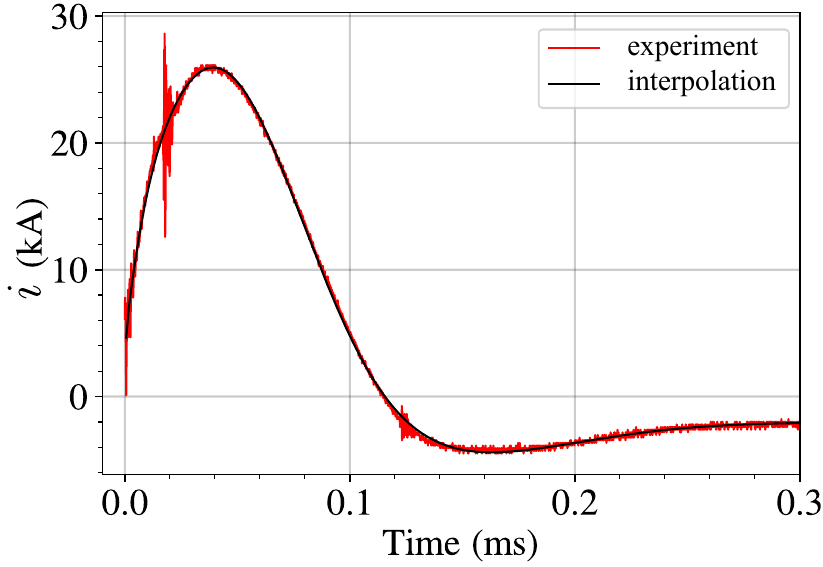}
         \caption{Discharge current (interpolated)}
         \label{fig:current_inter}
     \end{subfigure}
     \hfill
     \begin{subfigure}[b]{0.47\textwidth}
         \centering
         \includegraphics[width=0.95\textwidth]{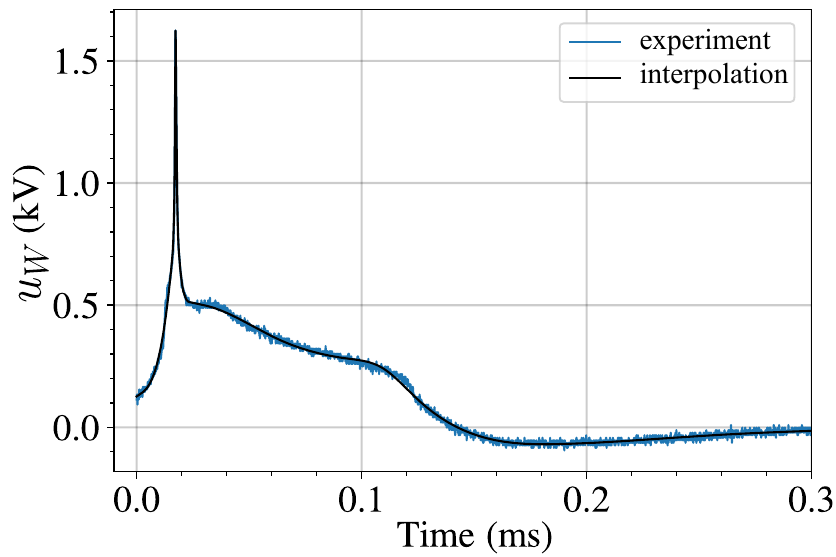}
         \caption{Discharge voltage (interpolated)}
         \label{fig:voltage_inter}
     \end{subfigure}
        \caption{Evolution of the discharge current (a) and voltage (b) measured at the wire extremities, obtained from  three repeated experiments. Comparison between the interpolated discharge current (c) and voltage (d) and the average of the time signal from the three experiments.}
        \label{fig:Curr_volt_w}
\end{figure}

\noindent Experiments were performed in miniBLAST to quantify the time evolution of the wire's resistance, $R_W$. Figure \ref{fig:Curr_volt_w} presents the measurements of the current and voltage transients at the wire's extremities and illustrates the excellent repeatability of our tests. For the sake of conciseness, we only present three experiments.
Both signals (current and voltage) are synchronized as shown by the green line which corresponds to the trigger signal, denoting the discharge initiation. It is worth noticing the presence of a small time delay of the order of $\Delta t \approx$ 9 \textmu s, required for the ignitron switch to close and initiate the discharge, cf. Figure \ref{fig:volt_w}. 

\noindent The experiments are found to be highly repeatable. Accordingly, in the following, we consider the mean of the three experimental results for the analysis of time evolution of the resistance evolution.

\begin{figure}[htp!]
     \centering
     \begin{subfigure}[b]{0.47\textwidth}
         \centering
         \includegraphics[width=1.\textwidth]{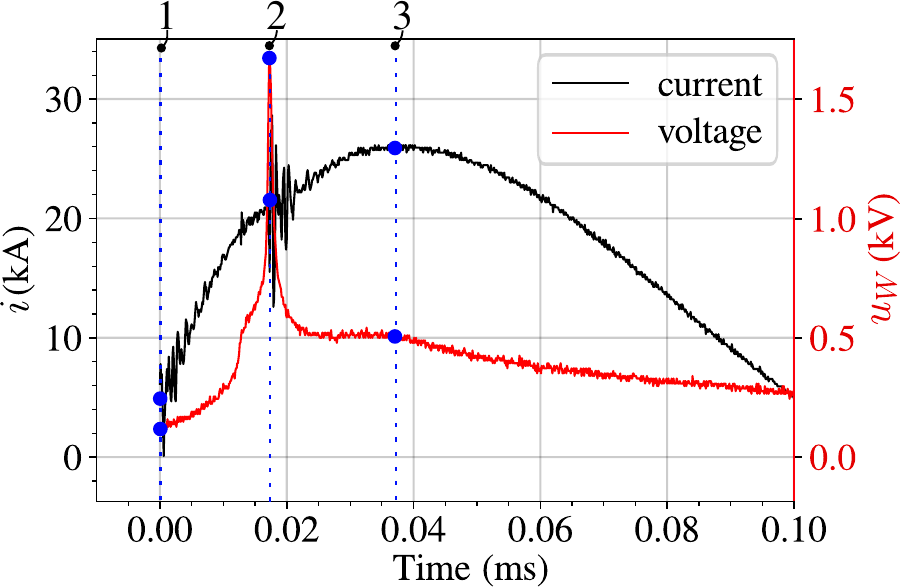}
         \caption{Time evolution of $i$ and $u_W$}
         \label{fig:w_data}
     \end{subfigure}
     \hfill
     \begin{subfigure}[b]{0.47\textwidth}
         \centering
         \includegraphics[width=0.65\textwidth]{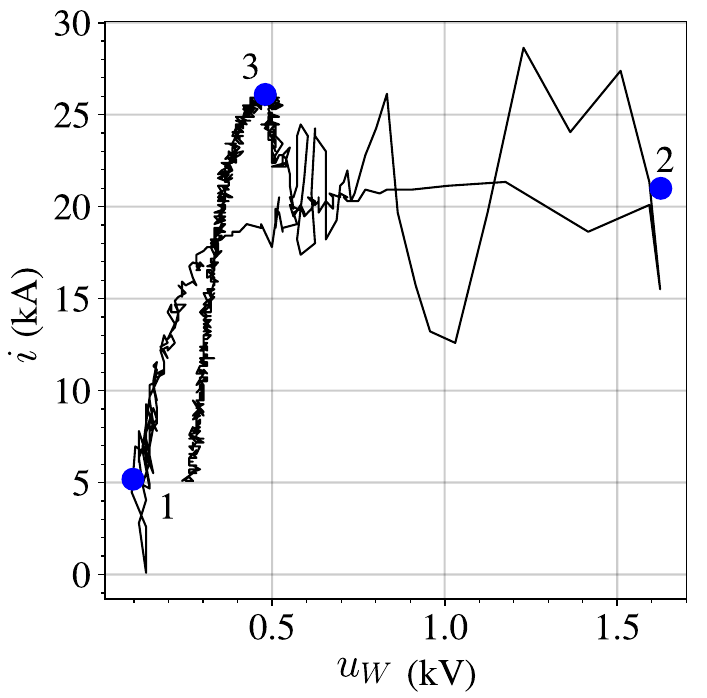}
         \caption{$i$ versus $u_W$}
         \label{fig:volt_w_curr}
     \end{subfigure}
     \hfill
     \begin{subfigure}[b]{0.47\textwidth}
         \centering
         \includegraphics[width=1.\textwidth]{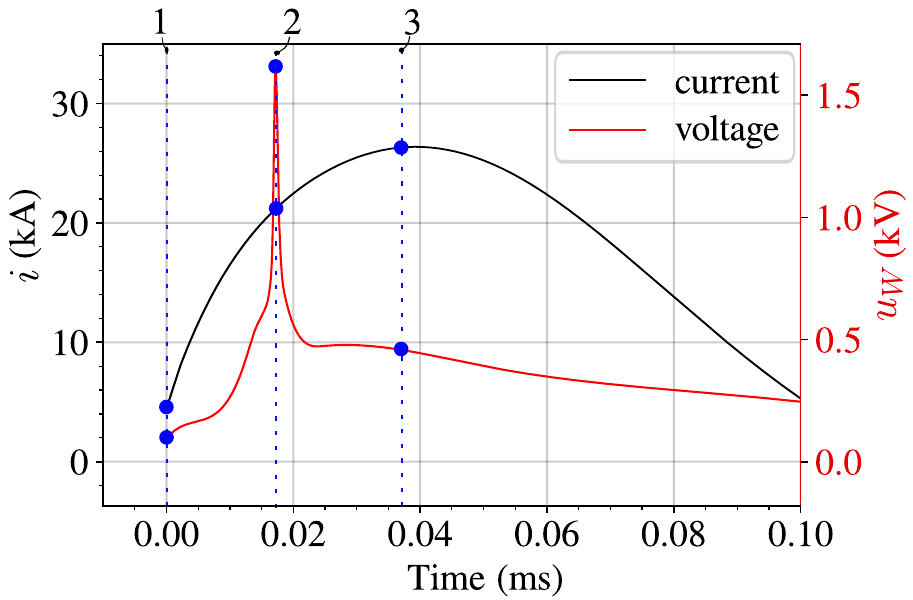}
         \caption{Time evolution of $i$ and $u_W$ (interpolated)}
         \label{fig:IV_signal}
     \end{subfigure}
     \hfill
     \begin{subfigure}[b]{0.47\textwidth}
         \centering
         \includegraphics[width=0.65\textwidth]{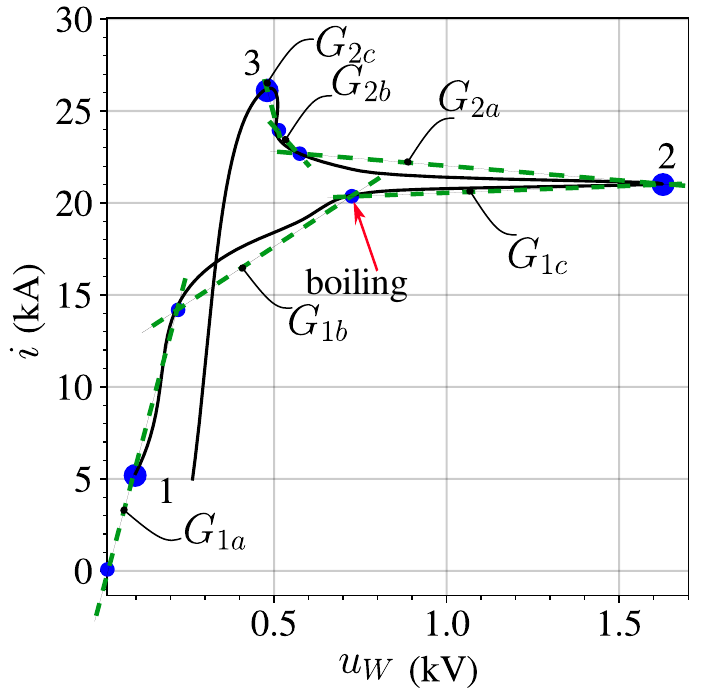}
         \caption{$i$ versus $u_W$ (interpolated)}
         \label{fig:I_V}
     \end{subfigure}
        \caption{Evolution of the discharge current and voltage: (a, c) time evolution and (b, d) $i$ versus $u_W$. The first row (a-b) refers to measured raw signals, while the second row (c-d) shows the interpolated ones.}
        \label{fig:w_Curr_volt}
\end{figure}

In Figure \ref{fig:w_data}, we plot the current and the voltage as function of time. Three points can be identified: point (1) corresponds to the initiation of the discharge; point (2) denotes the voltage peak; and point (3) corresponds to the current peak. These temporal variations can be linked to the stages previously explained in Section \ref{sec:Physical_processes}. Melting takes place between points (1) and (2) -- see Figure \ref{fig:exploding_mechanism}(b). Point (2) is followed by the electrical breakdown and the formation of plasma -- Figure \ref{fig:exploding_mechanism}(c). The latter follows the increase of the current flow that reaches its peak at point (3). The creation of the shock wave happens between points (2) and (3).

Figure \ref{fig:volt_w_curr} presents the current, $i$, as function of the voltage, $u_W$. The slope of the curve represents the conductance $G$ of the wire, while the resistance is the inverse of this slope, $R_W = 1/G$. 

Note that the phenomena taking place during stage 2, between points (1) and (2), cannot be captured with the sampling rate of our measuring devices, despite the latter being relatively high, i.e., $1\times 10^6$ samples per second. In addition, the noise-to-signal ratio is not negligible. For these reasons, we opt to interpolate both signals (current and voltage) with a one-dimensional cubic spline (for more details see \cite{morsel2024fast}). The interpolated current and voltage are presented in Figure \ref{fig:current_inter}, \ref{fig:voltage_inter}, and the resulting smoothed $i$-versus-$u_W$ plot is shown in Figure \ref{fig:I_V}.\\

As already discussed in Section \ref{sec:Physical_processes}, the conductance increases until the melting stage, stage S2. At point (1) $G_1a = 63.6$ A/V ($R_{W1a} = 0.0160$ $\Omega$). Then, the conductance decreases (resistance increase) reaching $G_{1b} = 12.8$ A/V ($R_{W1b} = 0.0780$ $\Omega$) and, later, $G_{1c} = 0.7$ A/V ($R_{W1e} = 1.4970$ $\Omega$). Between points (2) and (3), there is a substantial increase of the conductance due to the electrical breakdown (stage S3). This change is reflected in the conductivity values: $G_{2a} = -1.2$ A/V ($R_{W2a} = -0.8060$ $\Omega$), $G_{2b} = -23.3$ A/V ($R_{W2b} = -0.0428$ $\Omega$), and $G_{2c} = -80.0$ A/V ($R_{W2c} = -0.0120$ $\Omega$). This stage is characterized by a significant increase in the current flow, as shown in Figure \ref{fig:w_data}.

\noindent More specifically, the resistance of the wire can be calculated as
\begin{equation}
    R_W(t) = \frac{du_W(t)}{di(t)},
\label{eq:rsistance_dynamic}
\end{equation}
which is computed using the interpolated data of Figure \ref{fig:w_data}.

According to \cite{liu2019experimental, desai1984electrical, richardson2014experimental}, the resistivity of an aluminum wire at $T=20^\circ C$ is $\rho_s = 2.82\times 10^{-8}$ $\Omega \cdot \text{m}$. By further assuming a uniform resistivity along and across the aluminum wire (valid for a thin wire), the resistance of the resistor can be expressed in function of the material resistivity, $\rho_s$, as follows
\begin{equation}
    R = \rho_s \frac{l}{S},
\label{eq:resistivity}
\end{equation}
where $S$ is the cross sectional area of the resistor, assumed constant, and $l$ its length.\\
The initial resistance is found to be equal to $R_W = 0.0036 \Omega$. This value was confirmed using an ohmmeter (M210) to measure the wire's resistance. It is worth mentioning that the resistance measured at point (1), denoted as $R_{W1}$, is equal to $0.0025$ $\Omega$, which is close to the value measured by the ohmmeter. The small difference is attributed to the fact that we do not have access to the measurement of the current and the voltage at $t=0$ due to important noise at the beginning of the signals.

\begin{figure}[htp!]
    \centering
    \includegraphics[scale=0.5]{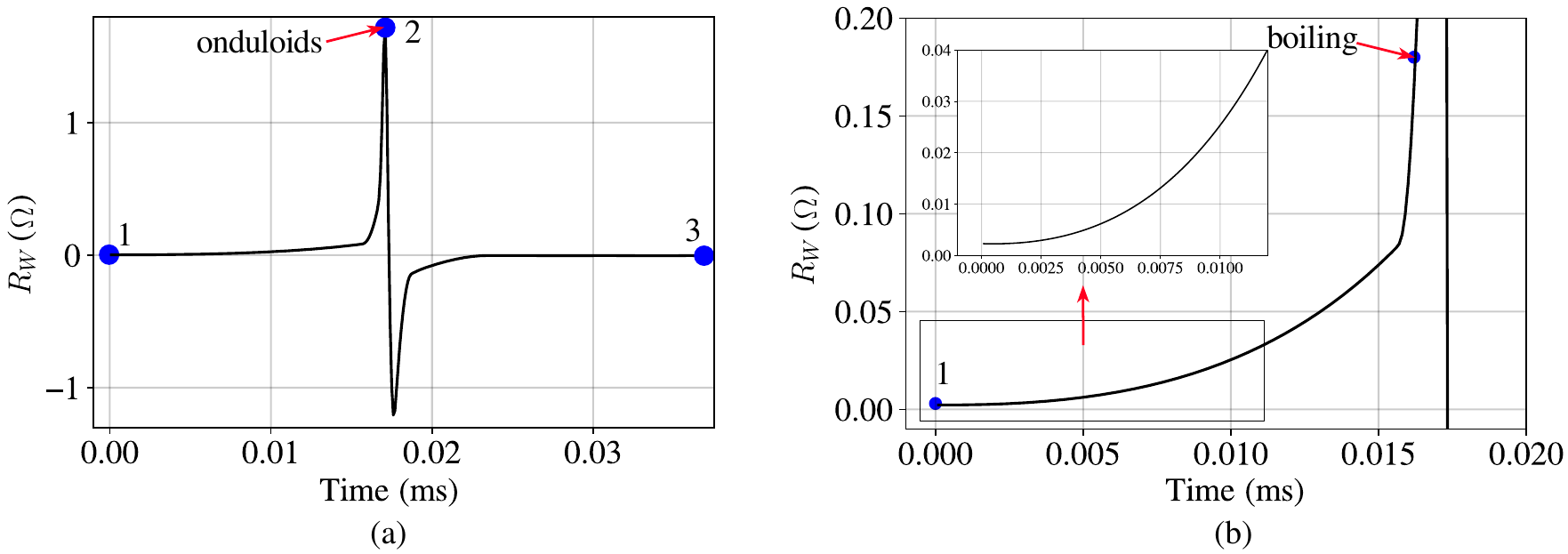}
    \caption{Time evolution of the aluminum wire resistance between points 1 and 3 (a), and zoom in on the period between 0 and 0.02 ms (b).}
    \label{fig:resistance}
\end{figure}

Figure \ref{fig:resistance} presents the time evolution of the calculated aluminum wire resistance. As expected, an increase of the resistance occurs from point (1) to (2), reaching a peak at point (2) -- $R_{W2}= 1.7260$ $\Omega$. The resistance then decreases from point (2) to (3), vanishing at point (3) which corresponds to the electrical breakdown (stage S3), Figure \ref{fig:resistance}.

\noindent The negative resistance observed in Figure \ref{fig:resistance} is believed to result from the highly non-linear phenomena that occur between points (2) and (3). In such cases, the relationship between voltage and current can become non-ohmic. In other words, as the current increases, the voltage decreases, leading to a negative (apparent) resistance (see also Figure \ref{fig:w_data}). Additionally, one should note that the considered system is open, allowing the energy from the magnetic field and the plasma zone around the wire to influence the system, resulting in an apparent negative resistance.

\newpage
\subsubsection{Power, energy and temperature}
\label{subsec: Power, energy and temperature}

\noindent Assuming constant resistance, the dissipated power, $P_R$, and energy, $E_R$, in the aluminum wire can be estimated using the following relations 
\begin{equation}
    P_R (t) =  u_R(t)i(t) = i^2(t)R = \frac{u^2_R (t)}{R},
\label{eq:power_dissp}
\end{equation}
\begin{equation}
    E_R (t) = \int_0^t P_R (t) \,dt = R\int_0^t i^2(t) \,dt = \frac{1}{R}\int_0^t u^2_R (t) \,dt.
\label{eq:Energy_dissp}
\end{equation}

\begin{figure}[htp]
     \centering
     \begin{subfigure}[b]{0.47\textwidth}
         \centering
         \includegraphics[width=0.9\textwidth]{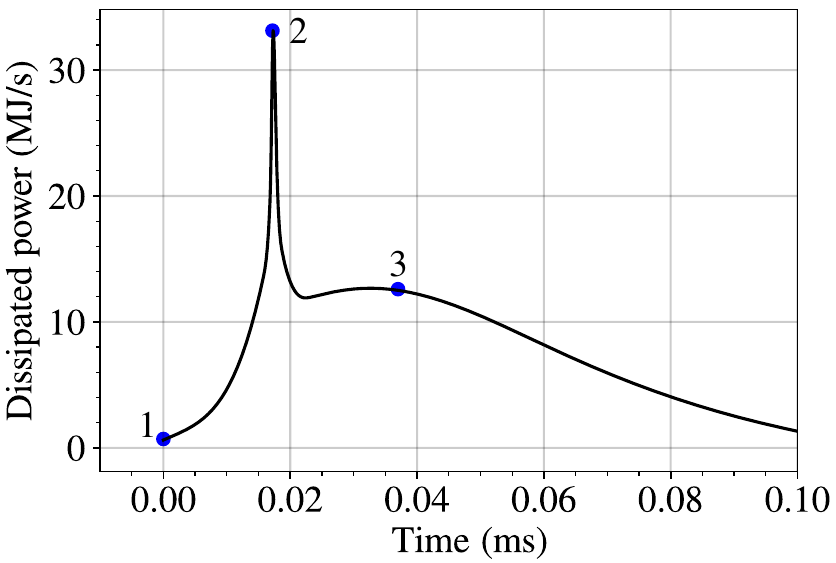}
         \caption{Dissipated power (MJ/s)}
         \label{fig:Power_diss}
     \end{subfigure}
     \hfill
     \begin{subfigure}[b]{0.47\textwidth}
         \centering
         \includegraphics[width=0.9\textwidth]{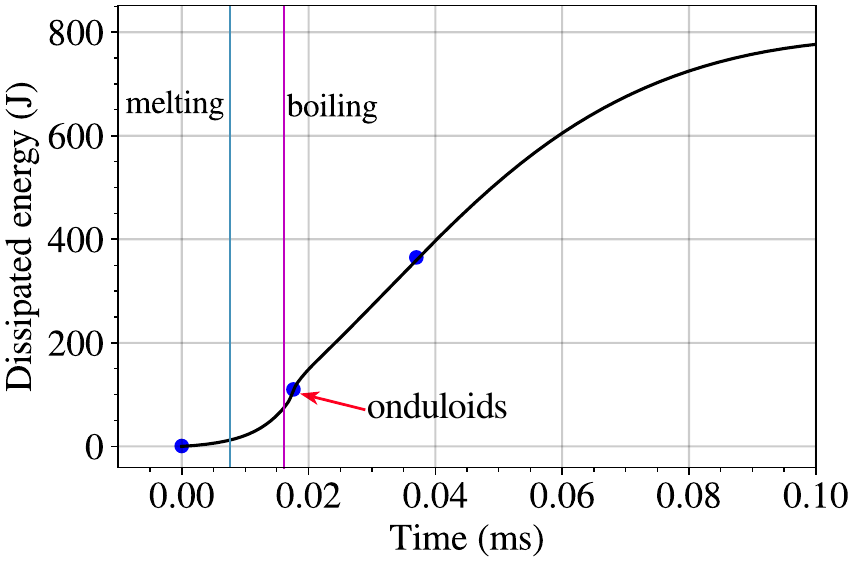}
         \caption{Dissipated energy (J)}
         \label{fig:Energy_diss}
     \end{subfigure}
     \hfill
     \begin{subfigure}[b]{1\textwidth}
         \centering
         \includegraphics[scale=0.5]{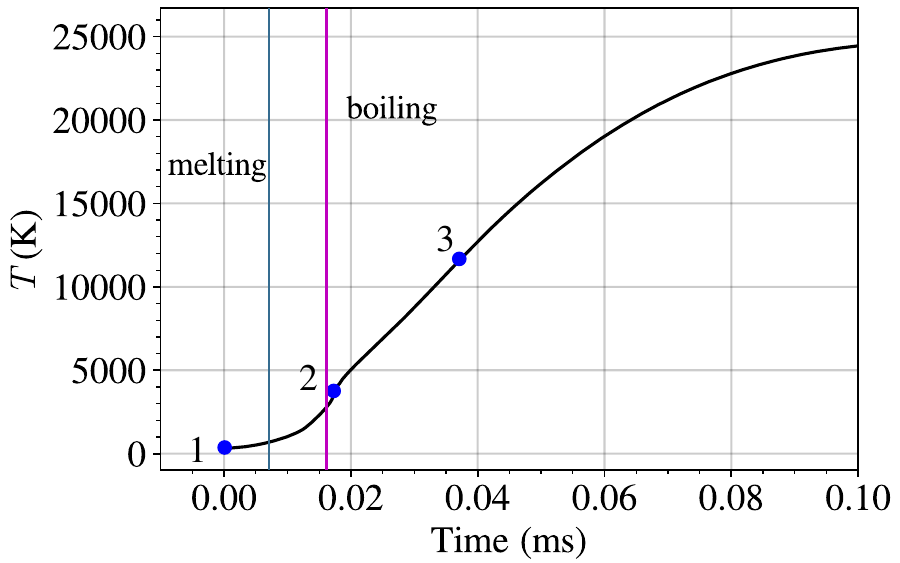}
         \caption{Temperature (K)}
         \label{fig:wire_temp}
     \end{subfigure}
        \caption{Time evolution of the dissipated power (a), dissipated energy (b), and temperature (c) in the aluminum wire.}
        \label{fig:Energy_power}
\end{figure}

Figure \ref{fig:Energy_power} illustrates the time evolution of both dissipated power and energy. The wire begins to melt around $t \approx 0.005$ ms (blue line), and the boiling phase starts at $t = 0.0162$ ms (violet line), as shown in Figure \ref{fig:Energy_diss}. Additionally, between points (1) and (2), the wire transitions into its melting phase, which is followed by the boiling stage and the formation of onduloids at the voltage peak. At point 2, which is associated with the voltage peak, the conductivity decreases rapidly while the dissipated energy increases.

Finally, Figure \ref{fig:wire_temp} presents an estimation of the time evolution of the temperature according to the calculations of Section \ref{Dissipated energy}. Aluminum has a melting temperature of approximately $660 \, ^\circ$C (933.15 K), while its boiling temperature is $2470 \, ^\circ$C (2743.15 K). At point (3), the temperature is expected to reach 12000 K. However, the experimental verification of these estimations exceeds the scope of the present work.

\newpage
\section{Sensors and metrology}
\label{sec:system_measurements}
\noindent Herein, we detail the sensors and the data acquisition devices used to measure pressure signals and the structural response. The different sources of uncertainties and errors associated with each component are also discussed.

\subsection{Data acquisition}
\label{Data acquisition devices}
\noindent A data acquisition system is responsible for capturing and digitizing analog signals from the various sensors. Two data acquisition devices are used, the Oscilloscope Nicolet Sigma 30 (ONS) and the Data Acquisition System (DAS) TraNET FE 404. 

\subsubsection{Nicolet Sigma 30 oscilloscope}
\noindent Oscilloscopes enable recording and analyzing of electrical waveforms. They allow for the precise examination of the transients of the signal. The ONS disposes a trigger used to synchronize events and recording data with a high sampling rate for the needs of our experiments (10 MS/s). It therefore allows us to measure current and voltage signals with a temporal resolutions of 0.1 \textmu s. The ONS is utilized for collecting data from the current transducers and voltage probes (see Section \ref{subsec:sensor}). It is equipped with four channels, with a storage capacity up to $1$ GB. It presents a DC offset of $\pm$ 0.25$\%$ and an amplitude resolution of 12 bits.

\subsubsection{TraNET FE 404 data acquisition system}

\noindent The data acquisition system must operate with sufficiently high speed to capture the pressure signals due to the shock waves. For the most critical scenario, the shock waves are expected to exhibit a positive time duration $t_o$ of $0.01$ ms. To ensure the recording of the shock wave's peak, $P_{so}$, it is necessary to sample a large amount of data points within this time frame (at least 200). Consequently, the data acquisition system should feature a sampling speed of approximately $20$ MS/s. Furthermore, the acquisition system must be capable of providing functional voltage to the sensors.

The DAS TraNET FE 404 data acquisition system meets these requirements, has 8 channels with a maximum sampling rate of 20 MS/s and an amplitude resolution of 16 bits up to 5 MHz and 14 bits up to 20 MHz. It has a range error of $\pm$ 0.03 $\%$ (typical) to 0.1 $\%$ (maximum), a DC offset of $\pm$ 0.03 $\%$ (typical) to $\pm$ 0.1 $\%$ (maximum), and an input noise of $\pm$ 3.57 mV, which are satisfactory for our measurements.

\subsection{Sensors}
\label{subsec:sensor}

\noindent In order to study the response of structures submitted to blast loads, we must measure the impinging pressure on the structure, as well as the current and voltage discharge within the electrical system. The miniBLAST incorporates a variety of sensors, including pressure transducers, current transducers and voltage probes. Integrated Electronics Piezoelectric (IEPE) transducers are selected as they eliminate the uncertainties and losses related to the length of the connecting cables and the triboelectric effect.

\subsubsection{Pressure transducers}
\label{subsubsec:sensor}

\noindent Pressure transducers are necessary to accurately characterize the blast load and its time signature. They are positioned parallel to the shock wave's propagation direction to measure the reflected overpressure and perpendicular to the shock wave's propagation to measure the incident overpressure.

\paragraph{Sensors used for measuring the reflected overpressure} The Kistler 603CBA family transducers can measure a wide range of pressures spanning from 1400 kPa to 10000 kPa. They operate at temperatures to $120^\circ$C, their high natural frequency ensures a very fast rise time ($<$0.4 \textmu s). These transducers are compact in size and provide voltage as the output signal \citep{Kistler}. As such, they are well-suited for the needs of this setup. The transducers exhibit a DC offset error varying between $\pm$ 0.320 kPa and $\pm$ 1.600 kPa and a linearity error between $\pm$ 0.05\% and $\pm$ 0.18\% of the full-scale output, corresponding to a range between 2000 kPa and 7000 kPa, respectively. Additionally, they have an acceleration sensitivity resulting in an error of $\pm$ 0.015 kPa. Consequently, the total cumulative error falls between depending the sensor 1.018 kPa and 15.223 kPa. This level of accuracy is satisfactory for our applications, as the pressures to be measured is of the order of many kPa.\\

\paragraph{Sensors used for measuring the incident overpressure} A Kistler 6233A0050 pencil probe is used to measure the incident overpressure due to its low rise time ($<$1 \textmu s), high sensitivity of 1400 mV/kPa, wide pressure measurement range up to 200 kPa and broad operating temperature range, from $-55^{\circ}$C to $125^{\circ}$C. Its acceleration sensitivity is $\pm$ 0.20 kPa/g. The linearity error is within $\pm \, 0.58\%$ of the full-scale output (200 kPa) resulting in an error of $\pm \, 1.16$ kPa. Therefore, the total cumulative error of the pencil probe is $\pm$ 1.36 kPa, which is again satisfactory for our applications.

\subsubsection{Current sensor}

\noindent Current sensors play an important role in real-time monitoring of electrical currents in circuits and systems, and in our case, they are used to analyze the detonation of exploding wires. The criteria guiding the choice of the current sensors are: (1) non-invasive measurement capability (no direct electrical contact with the current-carrying conductor) and (2) suitable frequency range.

The Rogowski current waveform transducer model CWT03-Ltd meets these criteria, with a measurement range from 300 mA to 300 kA and a negligible rise time. The sensor converts high currents into voltage with a sensitivity of 10 mV/kA. The DC offset remains below $\pm$ 0.2 kA within an operating temperature range from -20$^{\circ}$C to 100$^{\circ}$C. Its linearity error is within $\pm$ 0.05$\%$ of the full-scale output (300 kA), resulting in an error of $\pm$ $0.15$ kA. The positional accuracy is $\pm$ 0.5$\%$ of the measured value. Therefore, for a peak current of approximately $50$ kA (see also section \ref{sec:exploding_wires}), the total cumulative error is $\pm \, 0.6$ kA, which is satisfactory for our needs.

\subsubsection{Voltage probe}

\noindent Voltage measurements are necessary to study the electrical circuits parameters and to determine the evolution of the resistance of the explosive wire (see Section \ref{sec:exploding_wires}). The voltage probe should be able to accurately measure voltage up to 20 kV for the wide range of experiments we want to perform.

The Tektronix P6015A High Voltage Probe (HPV) is used as it offers a wide range of input voltage capabilities up to 20 kV for both direct current and alternating current measurements. The voltage sensor features a rapid rise time of 4.67 ns and provides a 1000x attenuation, with a sensitivity of 1 V/1 kV. The linearity error is within $\pm \, 0.5 \%$ (0.1 kV), the temperature error $\pm \, 0.006 \%$/$^{\circ}$C (where a temperature rise of 60$^{\circ}$C at 20 kV leads to an error equal to 3.6 V). The DC offset is $\pm \, 0.018 \%$/kV, i.e., at 20 kV the DC offset is 3.6 V. The total cumulative error at $20$ kV is about $\pm$ $0.11$ kV, which is well-suited for our experiments.

\subsection{Optical cameras and stereometry}

\noindent The response of structures to explosions can be hardly investigated experimentally using only sensors and accelerometers. Image techniques are highly effective for measuring two- and three-dimensional (2D and 3D) displacements. In so doing, we use GoPro cameras which are affordable and provide sufficient resolution (27.13 megapixels) and frame rate (240 fps) for the expected duration of the structural response of the tested models. Appropriate corrections for distortion are accounted for to avoid systematic errors.

To capture the structural response, we rely on Particle Tracking Velocimetry (PTV) in 3D, specifically using stereometry. We employ the software TEMA \cite{ImageSystems} which, using two or more cameras to capture images of the field from different viewpoints, can reconstruct the 3D field. The cameras are mounted on the optical table with a lighting system consisting of two LED spotlights to ensure adequate illumination of the model. In our experiments we can, thus, determine the evolution of the 3D displacements of different points of the structure by placing circular targets (stickers with diameter equal to $5.0$ mm). Example of the measurements of the structural response are given in Section \ref{sec:structural_response}.

\section{Example of experimental results}
\label{sec:First experimental results}

\noindent The capability of the developed experimental setup is showcased by producing controlled and reproducible explosions in the laboratory and by quantitatively measuring both the resulting blast loading and the dynamic response of a reduced-scale structure.

\subsection{Quantitative measurement of blast loads}
\label{Blast loads measurements}

\begin{figure}[htb!]
     \centering
     \begin{subfigure}[b]{0.47\textwidth}
         \centering
         \includegraphics[width=0.95\textwidth]{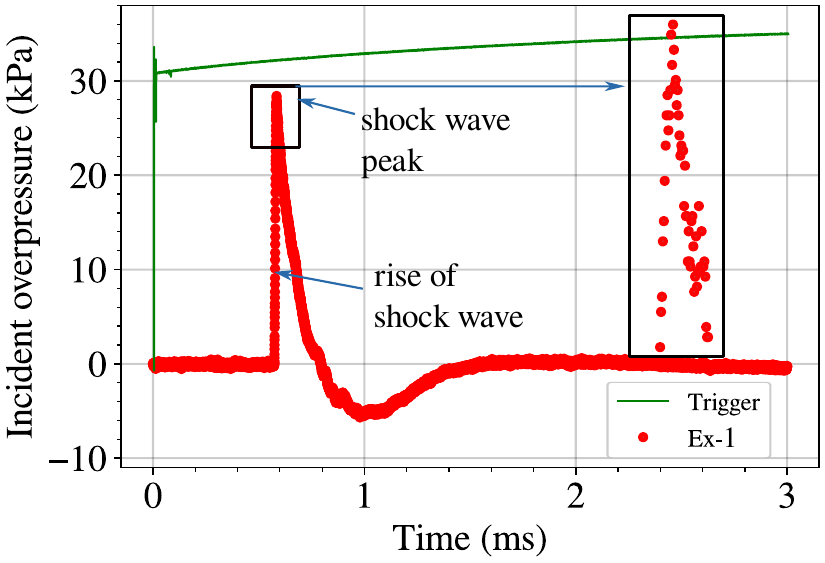}
         \caption{Incident overpressure}
         \label{fig:press_check_2}
     \end{subfigure}
     \begin{subfigure}[b]{0.47\textwidth}
         \centering
         \includegraphics[width=0.95\textwidth]{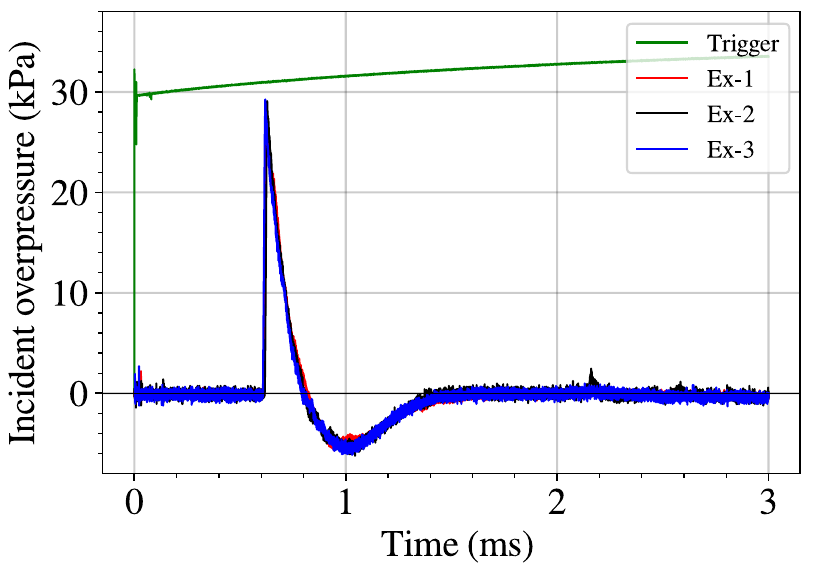}
         \caption{Incident overpressure for three identical tests}
         \label{fig:press_check_1}
     \end{subfigure}
        \caption{Time evolution of the incident overpressure, $P_s$, measured with sampling rate equal to 5 MS/s (a), and three different experiments (b) showing excellent repeatability.}
        \label{fig:press_check}
\end{figure}

In Figure \ref{fig:press_check} we depict the time history of the incident overpressure, $P_s$, measured using the pencil probe positioned at 30 cm from the explosive source, for a discharge load of 5 kJ.

\noindent Figure \ref{fig:press_check_2} shows the overpressure peak and the arrival time of the blast load, measured from the triggering point. The overpressure signal is recorded with a sampling rate equal to 5 MS/s, which is enough for capturing the overpressure peak with high accuracy in this example. The signature of the blast load agrees with that resulting from solid explosive detonations (cf. Figure \ref{fig:incident_ref_dynamic}). Figure \ref{fig:press_check_1} illustrates the signals recorded during three identical tests, which overlap, demonstrating excellent repeatability.

\subsection{Dynamic response of a reduced-scale structure due to blast load}
\label{sec:structural_response}
\noindent A reduced-scale, dry-stacked masonry wall is subjected to an explosion using two different stored energies: 2 kJ and 5 kJ. These experiments serve as a proof of concept of our setup for quantitative measurements of fast-dynamic, structural responses due to explosions.

\begin{figure}[htb!]
    \centering
    \includegraphics[scale=0.7]{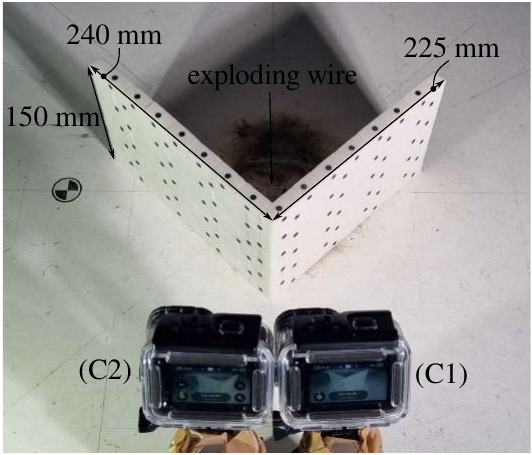}
    \caption{Setup of the structural test: dry-stacked masonry wall and cameras.}
    \label{fig:wall_edge}
\end{figure}

Figure \ref{fig:wall_edge} illustrates the geometry of the masonry wall, comprising 150 blocks. The blocks have dimensions: $30 \times 15 \times 15$ mm\textsuperscript{3}. Two cameras are used to track the 3D-displacement field, by means of PTV. Depending on the scaling ratios $\lambda$ and $\gamma$ the above energies of 2 kJ and 5 kJ could correspond to equivalent mass of TNT equal to tens of kg at real scale (for $\lambda$ equal to 1/20, and $gamma$ equal to 0.5).

\begin{figure}[htb!]
     \centering
     \begin{subfigure}[b]{1\textwidth}
         \centering
         \includegraphics[width=\textwidth]{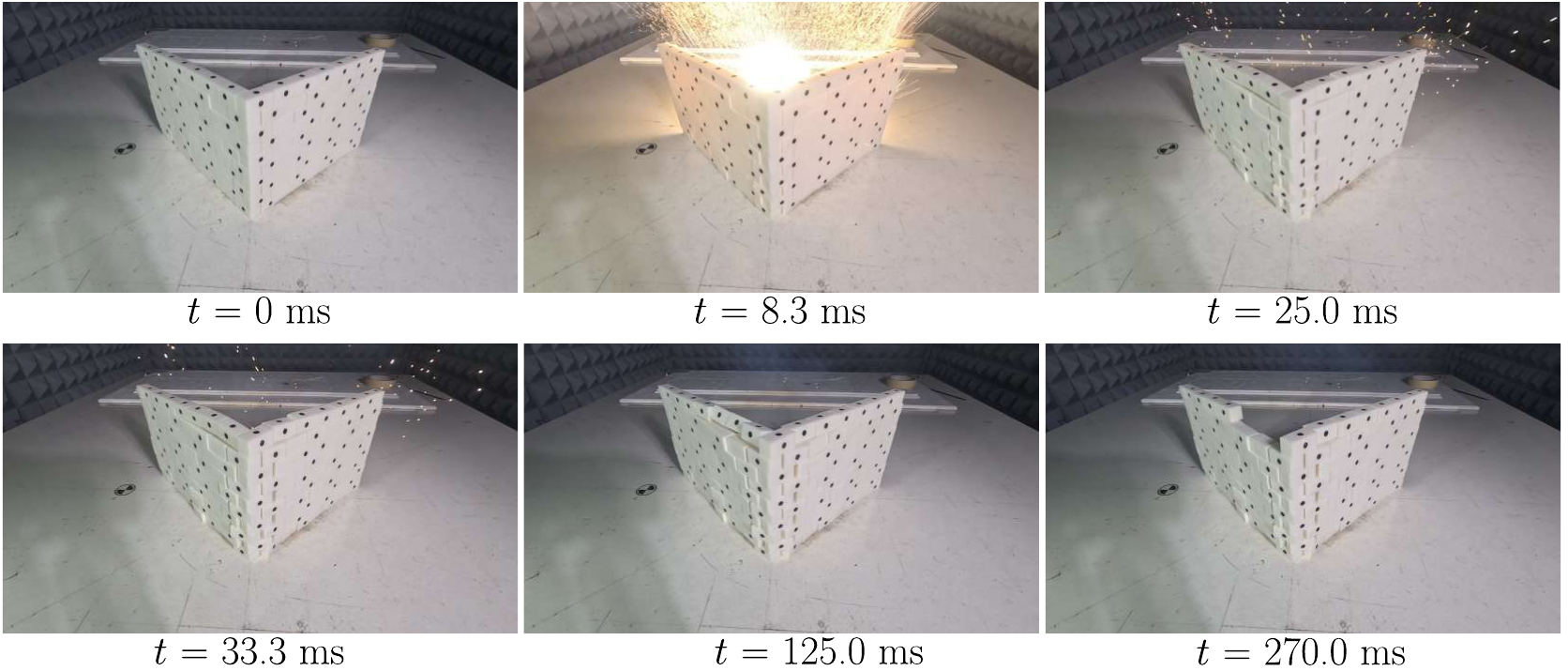}
         \caption{}
         \label{fig:wall_gopro1A}
     \end{subfigure}
     \\
     \vspace{10pt}
     \centering
     \begin{subfigure}{0.47\textwidth}
        \centering
        \includegraphics[height=0.5\linewidth]{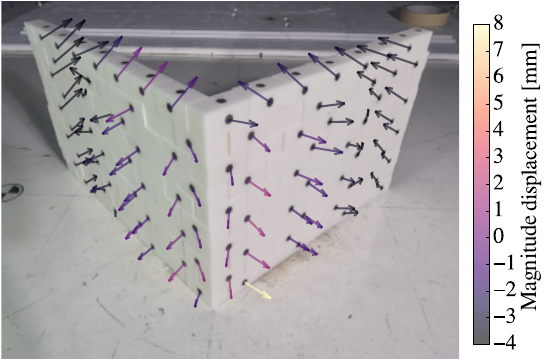}
        \caption{}
        \label{fig:3D_dalmation}
    \end{subfigure}
    \begin{subfigure}{0.47\textwidth}
        \centering
        \includegraphics[height=0.5\linewidth]{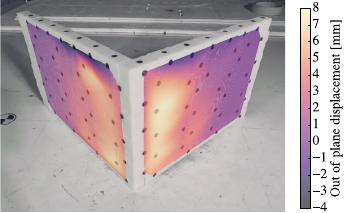}
        \caption{}
        \label{fig:out_of_plane_contours}
    \end{subfigure}
        \caption{Masonry wall model under blast load generated by a 2 kJ discharge: time evolution of blast (a), displacement field of the blocks (b), and contrours of the out of plane displacements of the wall.}
        \label{fig:Edge_wall_damage}
\end{figure}

Figure \ref{fig:wall_gopro1A} illustrates the time evolution of the structural response due to the blast load of 2 kJ. The wall does not collapse, but its response is characterized by significant out of plane displacements (\ref{fig:out_of_plane_contours}) and uplifting of the blocks positioned at the top layer of the masonry (\ref{fig:3D_dalmation}). Notice that close to the wall's corner the out of plane displacements are of the order of 8 mm, which is approximately equal to half of the blocks' thickness.

\begin{figure}[htb!]
     \centering
     \begin{subfigure}[b]{1\textwidth}
         \centering
         \includegraphics[width=\textwidth]{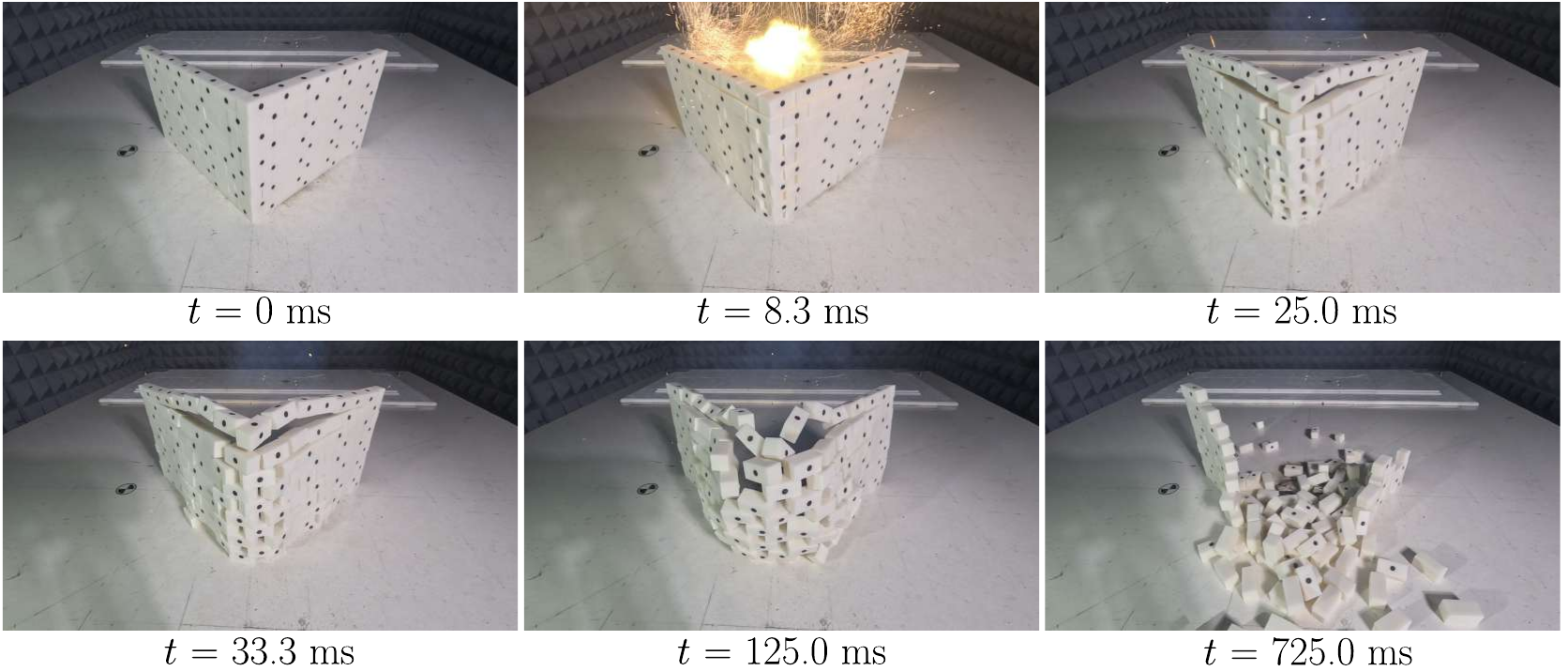}
         \caption{}
         \label{fig:wall_gopro1B}
     \end{subfigure}\\
     \vspace{10pt}
     \begin{subfigure}[b]{1\textwidth}
         \centering
         \includegraphics[width=\textwidth]{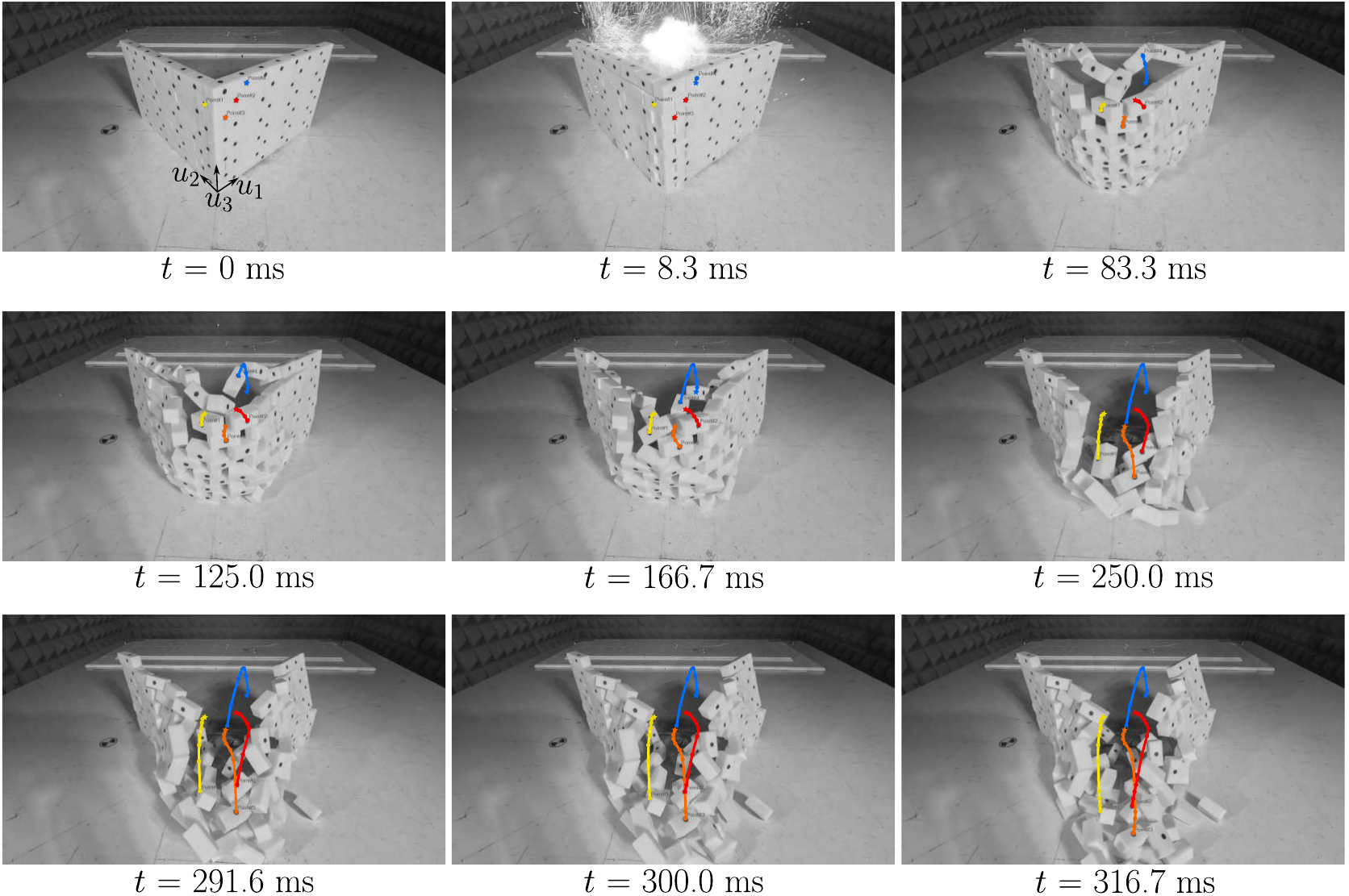}
         \caption{}
         \label{fig:Dalmation_P1}
     \end{subfigure}
        \caption{Discharge of 5 kJ and collapse of the masonry: time evolution of the blast event (a) and trajectory tracking of 4 blocks (b).}
        \label{fig:Edge_wall_damage5kJ}
\end{figure}

In Figure \ref{fig:Dalmation_P1} we present time instances of the dynamic response of the structure for the scenario of 5 kJ. In this case the wall collapses. Moreover, we present the trajectories of four different blocks during the event to demonstrate the capability of extracting detailed three-dimensional data.

The previous examples demonstrate the potential of this new setup to study the dynamics of structures under blast loads in the laboratory. The in depth study of the structural response of other structures and of the shock wave will follow in future works.

\section{Conclusions}
\label{Conclusion}

A novel experimental setup, called miniBLAST, was developed to study the dynamic response of structures subjected to blast loads in the laboratory. In this work, we provide a detailed description of the design rationale and technical characteristics of the setup, with emphasis on the installation phases, safety, and metrology.

Particular focus was given on the explosive source, which is based on the discharge of significant electrical charges over a thin aluminum conductor. This discharge generates blast-type shock waves, which we briefly explore and analyze. The generated blast loads were repeatable, and their intensity was controllable with high precision.

Finally, we presented examples of testing and measuring the three-dimensional dynamic response of structures under blast loads. More specifically, we tested a masonry wall under two levels of explosive intensity. In the first case, significant deformations were measured, while in the second case of higher intensity, we monitored the wall's collapse. We precisely tracked the velocities and trajectories of the structure's constituents with high time resolution in three dimensions.

This new experimental setup offers new perspectives in the study of fast structural dynamics due to blast loads, with reduced costs, increased safety, and repeatability of tests. By applying tailor-made scaling laws, the experimental results can be upscaled to real structures, providing valuable information about their response and weaknesses. Additionally, the results can be used to explore the performance, robustness, and accuracy of numerical models for structural dynamic analysis. This work lays the foundation for broader investigations and provides new tools to study the effects of blasts on structures and their mitigation.

\subsection*{Acknowledgments}
The authors would like to acknowledge the support of the Region Pays de la Loire and Nantes Métropole under the Connect Talent programme (CEEV: Controlling Extreme EVents - BLAST: Blast LoAds on STructures).


\bibliography{cas-refs} 
\end{document}